\documentclass{emulateapj}
\newcommand{\myemail}{bsiana@ipac.caltech.edu}

\newcommand{\leqsim}{\,\raisebox{-0.6ex}{$\buildrel < \over \sim$}\,}
\newcommand{\geqsim}{\,\raisebox{-0.6ex}{$\buildrel > \over \sim$}\,}

\usepackage{lscape}

\slugcomment{Accepted for pulication in the Astrophysical Journal}

\shorttitle{z=3 QSO Luminosity Function}
\shortauthors{Siana et al.}

\begin{document}

\title{High-Redshift QSOs in the SWIRE Survey and the $z\sim3$ QSO Luminosity Function$^{\dagger}$}

\author{Brian Siana\altaffilmark{1,2}, Maria del Carmen Polletta\altaffilmark{2}, Harding E. Smith\altaffilmark{2}, Carol J. Lonsdale\altaffilmark{3}, Eduardo Gonzalez-Solares\altaffilmark{4}, Duncan Farrah\altaffilmark{6},  Tom S. R. Babbedge\altaffilmark{5}, Michael Rowan-Robinson\altaffilmark{5}, Jason Surace\altaffilmark{1}, David Shupe\altaffilmark{1}, Fan Fang\altaffilmark{1}, Alberto Franceschini\altaffilmark{7}, Seb Oliver\altaffilmark{8}}

\altaffiltext{1}{$Spitzer$ Science Center, California Institute of Technology, 220-6, Pasadena, CA, 91125, USA}
\altaffiltext{2}{Center for Astrophysics \& Space Sciences, University of California, San Diego, CA, 92093-0424, USA}
\altaffiltext{3}{Infrared Processing \& Analysis Center, California Institute of Technology, 100-22, Pasadena, CA, 91125, USA}
\altaffiltext{4}{Institute of Astronomy, University of Cambridge, Madingley Road, Cambridge, CB3 0HA, UK}
\altaffiltext{5}{Astrophysics Group, Blackett Laboratory, Imperial College, Prince Consort Road, London, SW7 2BW, UK}
\altaffiltext{6}{Department of Astronomy, Cornell University, Space Sciences Building, Ithaca, NY, 14853, USA} 
\altaffiltext{7}{Dipartimento di Astronomia, Universita di Padova, Vocolo Osservatorio 5, I-35122 Padua, Italy}
\altaffiltext{8}{Astronomy Center, University of Sussex, Falmer, Brighton, BN1 9QH, UK}

\altaffiltext{$^{\dagger}$}{Some of the data presented herein were obtained at the W.M. Keck Observatory, which is operated as a scientific partnership among the California Institute of Technology, the University of California and the National Aeronautics and Space Administration.  The Observatory was made possible by the generous financial support of the W.M. Keck Foundation.}

\email{\myemail}

\begin{abstract}
We use a simple optical/infrared (IR) photometric selection of high-redshift QSOs that identifies a Lyman Break in the optical photometry and requires a red IR color to distinguish QSOs from common interlopers.  The search yields 100 $z\sim3$ ($U$-dropout) QSO candidates with $19<r'<22$ over 11.7 deg$^2$ in the ELAIS-N1 (EN1) and ELAIS-N2 (EN2) fields of the {\it Spitzer} Wide-area Infrared Extragalactic (SWIRE) Legacy Survey.  The $z\sim3$ selection is reliable, with spectroscopic follow-up of 10 candidates confirming they are all QSOs at $2.83<z<3.44$.  We find that our $z\sim4$ ($g'$-dropout) sample suffers from both unreliability and incompleteness but present 7 previously unidentified QSOs at $3.50<z<3.89$.  Detailed simulations show our $z\sim3$ completeness to be $\sim80-90\%$ from $3.0<z<3.5$, significantly better than the $\sim30-80\%$ completeness of the SDSS at these redshifts.  The resulting luminosity function extends two magnitudes fainter than SDSS and has a faint end slope of $\beta=-1.42\pm0.15$, consistent with values measured at lower redshift.  Therefore, we see no evidence for evolution of the faint end slope of the QSO luminosity function.  Including the SDSS QSO sample, we have now directly measured the space density of QSOs responsible for $\sim70$\% of the QSO UV luminosity density at $z\sim3$. We derive a maximum rate of HI photoionization from QSOs at $z\sim3.2$, $\Gamma = 4.8\times10^{-13}$ s$^{-1}$, about half of the total rate inferred through studies of the Ly$\alpha$ forest.  Therefore, star-forming galaxies and QSOs must contribute comparably to the photoionization of HI in the intergalactic medium at $z\sim3$.  

\end{abstract}


\section{Introduction} 
The QSO luminosity function (QLF) is an observable constraint on models of galaxy formation and the corresponding growth of super-massive black holes (SMBHs) \citep[eg.][]{small92,kauffmann00,haiman00}.  These models are useful in interpreting observed phenomena, such as the relation between a galaxy's black hole mass and bulge luminosity \citep{kormendy95,magorrian98}, as well as inferring specifics of QSO activity, such as initial black hole mass functions, QSO light curves, and accretion rates \citep[see eg.][]{hopkins06}.  

In addition to galaxy formation models, the QLF can be used to derive the QSOs' contribution to HI and HeII reionization.   QSOs are responsible for HeII reionization at $z\sim3$ \citep{jakobsen94,reimers97,hogan97,sokasian02} and are presumed to have a neglible contribution to the HI reionization at $z\sim6$ \citep{madau99}.  However, both of these claims require assumptions about the faint end slope of the QLF, as this has not been measured accurately at $z>2$.  

The first QSO luminosity functions demonstrated a rapid increase in space densities toward higher redshift \citep{schmidt68,schmidt83} .  Deeper surveys, which primarily identified QSOs by their ``UV-excess''\footnote{QSOs were most easily identified as point sources with $U-B$ colors bluer than most stars.}, found that the faint end of the QLF was shallower than the bright end \citep{boyle88, heisler88, koo88, hartwick90}.  Recent large surveys, most notably the 2dF Quasar Redshift Survey \citep[2QZ,][]{boyle00,croom04} have found thousands of $z<2.5$ QSOs.  With these large samples, the QSOs have been placed into smaller bins in both luminosity and redshift, accurately constraining the shape and evolution of the QLF.  The data are typically fit to a broken power-law \citep{boyle88,pei95}

\begin{equation}
\Phi(L,z) = \frac{\Phi(L^*)/L^*}{(L/L^*)^{-\alpha}+(L/L^*)^{-\beta}},
\label{eqn:qlf_lum}
\end{equation}

{\noindent} with the break at $L^*$ and a bright end slope, $\alpha$, steeper than the faint end slope, $\beta$.  The QLF evolution with redshift is consistent with Pure Luminosity Evolution (PLE) \citep{marshall83,marshall85,boyle88,hartwick90,boyle00,croom04,richards05}.  That is, the evolution can be parametrized by a shift in the luminosity of the break, $L^*(z)$, without any change in its shape.  Recently, the 2dF-SDSS LRG and QSO Survey \citep[2SLAQ,][]{richards05} has extended the ``UV-excess'' QSO search one magnitude fainter to more accurately measure the faint end slope of the QLF out to $z=2.1$.  These deeper data, when combined with the bright end slope from the 2QZ and 6QZ \citep{croom04}, fit a faint-end slope, $\beta=-1.45$, and bright-end slope, $\alpha=-3.31$, and demonstrates a $\sim$40-fold increase in $L^*$ from $z=0$ to $z=2$. 

At high redshift, various surveys have been conducted through grism searches for UV emission lines \citep{schmidt95}, spectroscopic follow-up of point sources with optical colors away from the stellar main sequence \citep{warren94,kennefick95,fan01}, X-ray \citep{hasinger05}, searches for radio-loud \citep{dunlop90} or infrared luminous QSOs \citep{brown06}.  These surveys all show a precipitous decrease in space densities at $z>3$.  Early data from the Sloan Digital Sky Survey \citep[SDSS,][]{york00}, \citet{fan01} show a factor of six decrease in $M<-25.5$ QSOs from $z=3.5$ to $z=6.0$.  Recent results from SDSS suggest that the bright end slope is no longer constant at $z>3$.  Rather, it is getting shallower toward higher redshift (Richards et al. 2006). Unfortunately, these high redshift surveys are shallow ($i \lesssim 20$) and can only measure the bright end of the QLF at $z>3$.  Therefore, little can be said about the shape of the faint end of the high-z QLF, or its integrated properties (eg. contribution to intergalactic HI, HeII ionizing radiation or black hole growth) without a census of fainter QSOs at $z>3$.  

\citet{hunt04} (hereafter, H04) searched for faint AGN at $z\sim3$ with deep Keck spectroscopy over 0.43 deg$^2$ and found 11 QSOs.  Though limited by small numbers, the fitted faint-end slope, $\beta=-1.24 \pm 0.07$, is substantially shallower than low redshift measurements.  

X-ray selected samples suggest a rather modest evolution in AGN space density at high-z \citep{barger05}, or even a luminosity dependent density evolution (LDDE) where lower luminosity AGN peak in number density at lower redshifts than more luminous QSOs \citep{ueda03,hasinger05}.

Given the importance of the QSO luminosity function in constraining models of galaxy and black hole formation as well as the contribution of AGN to the ionizing background, these initial indications that the QLF shape is evolving at $z>3$ warrant deeper surveys to better constrain the high redshift QLF.  

The SWIRE Legacy Survey \citep{lonsdale03}, a wide-area infrared survey with deep ground-based optical data, is optimal for searches of faint QSOs at high redshift as it is deep enough to detect sub-$L^*$ QSOs at $z\leq4$ and covers sufficient area to detect large numbers of them.  In Section \ref{template} we outline our method for creating a new QSO template spanning far-UV to mid-IR wavelengths.  In Section \ref{selection}, we present a simple optical/IR color selection for QSOs at $z>2.8$ and identify areas of possible contaminations or incompleteness.  Our selection results are given in Section \ref{results}.  In Section \ref{reliability}, the reliability of the sample is assessed through spectroscopic follow-up and analysis of the infrared colors.  In Section \ref{completeness} we determine, through simulations and comparisons with known samples, our sample completeness as a function of redshift.  In Sections \ref{qlf} and \ref{qlf_comp} we present our measurement of the QLF at $z\sim3$ and compare to previous studies.  In Section \ref{ion_bkg}, the QSO contribution to photoionization of HI in the IGM is computed and compared to measurements of the {\it total} photoionization rate.  

Although many early studies of high redshift QLFs use an Einstein-DeSitter cosmology with $H_0=50 $km s$^{-1}$ Mpc$^{-1}$, throughout this paper we choose to use a more recent cosmology with $H_0=70$ km s$^{-1}$ Mpc$^{-1}$, $\Omega_m = 0.3$, $\Omega_{\Lambda}=0.7$ and correct other measurements accordingly.  All optical magnitudes are Vega magnitudes unless stated otherwise.
\section{Observations}

The SWIRE survey covers 49 deg$^2$ over six fields at high galactic latitude with minimum galactic cirrus emission.  Most of this area has now been imaged in multiple optical filters to depths $r' \lesssim 24.5$.  Our analysis is conducted within the first two fields for which both optical and infrared catalogs were available, ELAIS-N1 (16h11m+55$^\circ$00') and ELAIS-N2 (16h37+41$^\circ$02').  

\subsection{$Spitzer$ Infrared Data}
SWIRE is an IR imaging survey with all four bands on the Infrared Array Camera \citep[IRAC,][]{fazio04} and all three bands on the Multiband Imaging Photometer \citep[MIPS,][]{rieke04} aboard the {\it Spitzer Space Telescope} \citep{werner04}.  The IR filter characteristics and SWIRE depths are summarized in Table \ref{tab:ir_data}.  The EN1 IRAC campaign was undertaken January 14-20, 2004 and MIPS January 21-28, 2004.  MIPS went into standby mode on January 25, 2004, resulting in lost AORs which were reobserved July 29, 2004.  The EN2 IRAC campaign was observed July 05-06, 2004 and MIPS July 08-11, 2004. 

The IRAC and MIPS photometry were measured with the SExtractor program \citep{bertin96} within 1.9$''$ and 15$''$ diameter apertures for IRAC and MIPS 24$\mu m$, respectively \citep{swire_doc}.  Aperture corrections were derived from measurements of composite point spread functions from bright stars in the SWIRE fields. 

\subsection{Optical Data}
The EN1 and EN2 fields were imaged as part of the Wide Field Survey \citep[WFS,][]{mcmahon01}\footnote{http://www.ast.cam.ac.uk/~wfcsur/}.  Images were taken with the Wide Field Camera (WFC) on the 2.5-meter Isaac Newton Telescope (INT).   Both EN1 and EN2 have been observed with the $U$,$g'$,$r'$,$i'$,$Z$ filters over 9 deg$^2$ each, with 600 second exposures at each pointing in each filter.  A fraction of the fields $\sim 30$\% were not observed on the same night in every filter.  The filter characteristics and depths are summarized in Table \ref{tab:opt_data}.  The median seeing is $\sim$1.1$''$ and never worse than $1.6''$.  The optical coverage overlaps the {\it Spitzer} IR data by 7.45 and 4.29 deg$^2$ in EN1 and EN2, respectively \citep{swire_doc}.  Data processing was done by the Cambridge Astronomical Survey Unit CASU) and is outlined in \citet{irwin01} and \citet{gonzalez-solares05}.

Photometry was measured with the CASU software, requiring a source to have five continguous pixels $1.5\sigma$ above the background pixel noise.  Detection and photometry were performed in each band separately and then matched between bands.  Fluxes were measured within 2.3$''$ (7 pixels) diameter apertures.  Given the typical seeing of $\sim1.1''$, these apertures contain 80-90\% of the total flux.  Aperture corrections for each image were derived from bright stars within that image.  Total and isophotal magnitudes were computed as well but only aperture photometry is used in this analysis as our objects are point sources by definition.  Limiting magnitudes ($5\sigma$) for non-detections were computed from the pixel-to-pixel noise of the corresponding image.  The images were not interpolated before photometry was performed and therefore do not suffer from correlated noise from projection procedures.

\section{QSO Template}
\label{template}
Our optical QSO selection (defined in Section \ref{opt_sel}) uses three filters which typically span the rest-frame wavelengths 700 \AA\ $< \lambda <$ 2000 \AA.  For the infrared selection (see Section \ref{ir_sel}) we use the same two filters (IRAC1 \& IRAC2) for all targeted QSO redshifts, so the photometry samples a large range in the rest-frame optical to near-IR wavelengths, 5000 \AA\ $<\lambda<2 \mu$m.  In order to define the expected optical/mid-IR colors of QSOs at $2.7<z<5.0$, we have created a QSO template which spans many decades in wavelength from the far-UV to the mid-IR.

\subsection{UV Template}
Several composite QSO spectra have been created from large QSO surveys, and agree well with each other except for minor variations due to selection effects in optical color and luminosity.  \citet{vanden_berk01} created a composite spectrum of 2200 QSOs from the SDSS spanning the wavelengths between 800 \AA\ $< \lambda < 8555$ \AA.  In order to obtain such a broad coverage in rest-frame wavelength, QSOs from a wide redshift range (0.044 $\le z \le$ 4.789) were used.  Since the UV spectrum is composed from only high-redshift QSOs, the mean continuum shortward of Ly$\alpha$ is artificially decreased by the large number of Lyman line absorbers at high redshift.  Because the level of absorption is redshift dependent, we need a template which reflects the {\it intrinsic} SED, to which we can then apply a redshift dependent model of the HI absorption.

\citet{telfer02} have created a high S/N composite UV spectrum (300 \AA\ $< \lambda <$ 1275 \AA) using HST UV spectra of 77 radio quiet QSOs.  Because these data were taken in the observed UV rather than the optical, the QSOs contributing to the critical wavelength range 700  $< \lambda <$ 1216\AA\ are at lower redshift ($<z>=1.42$) than the SDSS QSOs (whose spectra cover the same rest-frame wavelengths at $z>2.4$).  Therefore, there is much less Lyman line and continuum absorption in this composite.  Furthermore, Telfer et al. (2002) corrected for absorbers with column densities $N_{HI} < 10^{16} $cm$^2$ and statistically corrected for lower column density absorbers, resulting in a good template of the intrinsic UV spectrum of QSOs.  For our UV template we have used the composite from \citet{telfer02} for 300 \AA\ $< \lambda <$ 1250 \AA\ and the SDSS composite for 2000 \AA\ $< \lambda <$ 8555 \AA.  The mean of the two composites is used in the overlapping regions 1250 \AA $<\lambda<$2000 \AA, where they agree well with each other \citep{telfer02}.  

\subsection{Optical to Mid-Infrared Template}
The SDSS composite from \citet{vanden_berk01} covers the optical to $\lambda <$ 8555 \AA.  However, at $\lambda >$ 5000 \AA\ the composite is produced by low-z, low luminosity AGN and suffers significant contamination from host galaxy stellar light.  Indeed, a comparison with the median broadband SED of luminous QSOs from \citet{elvis94}, for which stellar contribution to luminosity should be small, shows a much redder continuum slope in the SDSS composite at $\lambda > 5000$ \AA.  To reliably select QSOs at $2.5 < z < 6.0$ with our mid-IR photometry, a new template is needed which covers the rest-frame 5000 \AA\ $< \lambda <$ 2$ \mu$m and resembles the high luminosity ($M_{1450} < -23$) QSOs that we are targeting. 

We matched the SDSS photometry of all spectroscopically confirmed QSOs in the SDSS Data Release 3 \citep{schneider05} with the near-IR photometry from the Two Micron All Sky Survey \citep[2MASS,][]{skrutskie06} point-source catalog, including the 2MASS Deep Lockman Hole Catalog \citep{beichman03}, which goes $\sim 1$ magnitude deeper than the typical 2MASS all-sky depths.  There were 8642 2MASS objects within 2$''$ of the SDSS positions.  To extend the template to longer wavelengths, we also matched the {\it Spitzer} IR photometry of the 241 SDSS QSOs in the SWIRE fields (Lockman, EN1, \& EN2).

We chose to normalize the photometry at rest frame $\lambda_{norm} = 2400$ \AA\ for three reasons.  First, the $\lambda_{norm}$ is easily measured with our ground based optical/near-IR data for a broad range of redshifts, without using extrapolations and invoking assumptions about the SED spectral slope.  Second, the UV flux is dominated by the AGN rather than the host galaxy stellar light.  Third, this wavelength is far from any major emission lines which would significantly affect the photometry (eg, MgII or CIII]).  Only QSOs with detections in adjacent filters surrounding $\lambda_{rest} = 2400$ \AA\ were used to minimize the interpolation of the flux to the normalization wavelength.  After an $F_\nu$(2400 \AA) was calculated, only those QSOs with $M_{2400} < -24$ were used to ensure the flux was dominated by the AGN.  This also corresponds well with the minimum luminosity of QSOs in our search ($M_{1450} = -23.5$) and is $\sim$1 mag brighter than the classical QSO/Seyfert demarcation.  This cut in absolute magnitude left 3378 QSOs (39\% of SDSS/2MASS matches) with which we made the template.  

\begin{figure}
\epsscale{1.0}
\plotone{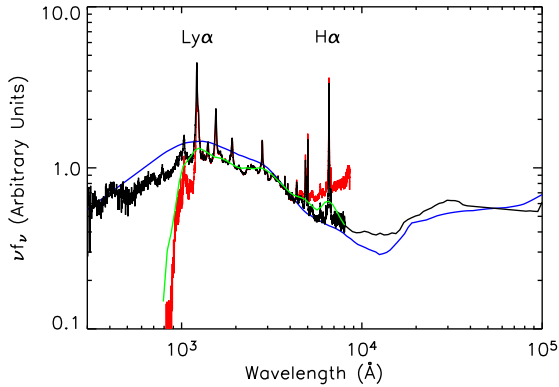}
\caption{The new optical/IR QSO template combined with \citet{telfer02} template (black).  The broadband template at $\lambda<8000$\AA\ has been replaced by the corrected Telfer/Vanden Berk template but is plotted in green to demonstrate the accuracy of the spectral slope.  Also plotted for comparison are the \citet{elvis94} template (blue) and the \citet{vanden_berk01} SDSS composite spectrum (red).  All templates are normalized at $\lambda=2900$\AA.  \label{fig:ir_template}}
\end{figure}

The template was made by averaging all of the flux measurements (SDSS/2MASS/SWIRE) in 1000 wavelength bins spaced logarithmically between 753 \AA\ and 11 $\mu$m.  The result is plotted in green in Figure \ref{fig:ir_template} and traces well the combined template of \citet{telfer02} and \citet{vanden_berk01} in the UV.  As expected, the relative UV to optical ratio of our template is significantly less than the \citet{vanden_berk01} template because our sample is subject to less contamination from the host galaxies.  

Broadband photometry works well in recreating smooth, continuous SEDs, but sharp features such as emission lines and continuum breaks are convolved with the filters through which they are observed, effectively broadening the features.  Deconvolution of this broadening is difficult since our photometry comes from 13 different filters (ugrizJHK,IRAC1-4,MIPS24).  The only sharp feature at $\lambda>5500$ \AA\ is the combined emission lines of H$\alpha$ (6563 \AA) \& [NII] (6548 \AA\ and 6592 \AA).  We therefore use the broadband photometry to get the shape of the underlying continuum at $\lambda >$ 5100 \AA, subtract the areas affected by H$\alpha$ \& [NII], and interpolate the continuum along this region.  We then add the H$\alpha$+[NII] profile from the \citet{vanden_berk01} template (scaled to the 2400 \AA\ continuum value to preserve the relative line strengths).

The resulting infrared QSO template is shown in Figure \ref{fig:ir_template} and tabulated in Table \ref{tab:template}.  An important feature of the template is the minimum at $\lambda \sim 1 \mu$m and the monotonic rise (in $\nu f_{\nu}$) toward longer wavelengths.  This is produced by hot dust at varying distances from the central engine \citep{barvainis87}.  The minimum at $\lambda \sim 1 \mu$m has been shown to be prevalent in QSO SEDs \citep{sanders89} and is attributed to sublimation of dust grains at T $\geqsim 1500$K \citep{barvainis87}.

Another important aspect of this new template is the increased equivalent width (W) of H$\alpha$+[NII].  \citet{vanden_berk01} derived a combined W(H$\alpha$+[NII]) = 197 \AA\ from their composite spectrum.  Since the relative stellar contribution has been removed in our template, we now have W(H$\alpha$+[NII]) $\sim 340$ \AA, more than 70\% higher.  This proves to be important at $z\sim4$ when H$\alpha$ redshifts into the mid-IR and significantly affects the IRAC colors.  

Finally, we point out that our broadband template is significantly redder than the \citet{elvis94} template, with the ratio of far-UV to near-IR nearly a factor of two larger in the latter.  This is not surprising since their sample was selected in the soft (0.3-2.0 keV) and ``ultrasoft'' (0.1-0.3 keV) X-ray bands.  Also, their sample is composed of more luminous QSOs, which are known to have bluer UV-optical colors than fainter QSOs \citep{richards06b}.

\section{High Redshift QSO Selection}
\label{selection}

In this section we define $z>3$ QSO selection criteria requiring only three optical bands and the two most sensitive imaging bands on the {\it Spitzer Space Telescope} (IRAC1 \& IRAC2) and assess its efficacy.  The method consists of an optical color selection to identify a Lyman Break in the rest-frame UV, thereby isolating QSOs to a narrow redshift range (ie. $z \sim 3$ for U-band dropouts).  In addition, we also require a red {\it Spitzer} IR color to eliminate typical contaminants in Lyman-Break Galaxy surveys (stars and low-z galaxies).  

\subsection{Optical Selection}
\label{opt_sel}
The space density of neutral Hydrogen (HI) absorbers increases rapidly with redshift \citep{bechtold94, weymann98}.  There are hundreds of absorbers with $N_{HI} > 10^{12} $cm$^{-2}$  along any line-of-sight (LOS) to galaxies with $z>2$, resulting in Lyman line absorption, or ``blanketing'', of the source's continuum at $\lambda_{rest}<1216$ \AA.  In addition, Lyman continuum absorption by the absorbers significantly decreases the observed flux at $\lambda_{rest}<912$ \AA.  At $z \geq2.5$, these features redshift into the bluest filter accessible to ground-based telescopes (U band) and can be identified by a flux decrement at wavelengths shortward of an otherwise blue continuum.  This method, known as the Lyman Break technique, has been used extensively to search for both QSOs and galaxies at $z>3$ \citep{irwin91,steidel96,madau96}.  

\begin{figure}
\epsscale{1.0}
\plotone{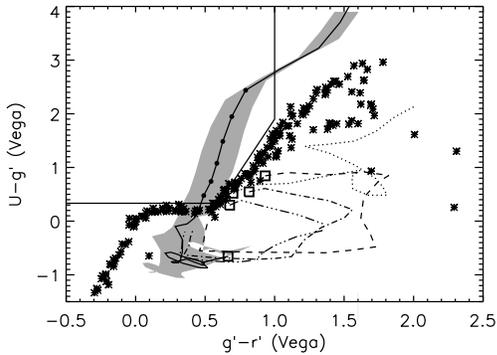}
\caption{$U-g'$, $g'-r'$ color-color diagram showing the $z\sim3$ QSO selection.  The solid curve is the color track of our QSO template with IGM absorption applied.  The black filled circles denote the locations in redshift increments of 0.1 from $2.9\le z \le 3.4$.  The shaded regions denote the color space spanned by QSOs with $\pm 2\sigma$ deviations in the spectral slope (see Table \ref{tab:sim_params}).  The lines are color tracks of various galaxy templates from $0<z<2$: Ell (age=2Gyr, dotted), Sa (dashed), Sc(dot-dashed), and Sd(multiple dot-dashed) taken from the GRASIL library of models \citep{silva98}, with the boxes corresponding to their respective colors at $z=0$.  The black asterisks are stars from the \citet{gunn83} catalog.  \label{fig:z3_opt_sel}}
\end{figure}

We use the \citet{madau95} prescription for the HI opacity evolution to determine average QSO colors as a function of redshift.  Figure \ref{fig:z3_opt_sel} shows the $U-g'$, $g'-r'$ color track of our our QSO template as it is redshifted and the HI absorption is applied.  At $z=2.3$ the Lyman Break redshifts into the $U$ filter, resulting in redder $U-g'$ colors, thereby causing the QSO to move away from the locus of blue, low-redshift QSOs.  At $z$\geqsim$2.8$ the QSO moves away from the optical color space of main sequence stars and into color space unoccupied by typical stars or galaxies.  We can therefore search for QSOs in this color space while minimizing contamination.  

\begin{figure}
\epsscale{1.0}
\plotone{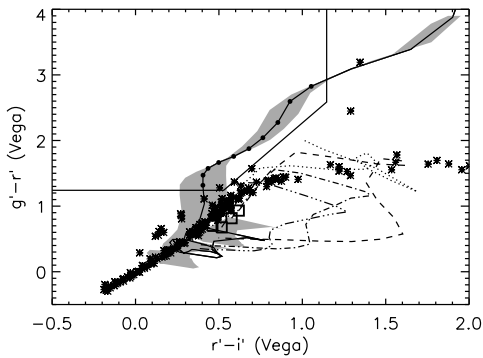}
\caption{$g'-r'$, $r'-i'$ color-color diagram showing the $z\sim4$ QSO selection.  The symbols and lines are the same as in Figure \ref{fig:z3_opt_sel}. \label{fig:z4_opt_sel}}
\end{figure}

Our color selection for both the $z\sim3$ and $z\sim4$ sample, defined in Table \ref{tab:opt_sel} and shown in Figures \ref{fig:z3_opt_sel} and \ref{fig:z4_opt_sel} were defined to select QSOs with $\leqsim 2\sigma$ deviation in spectral slope (see Section \ref{sim}).  As we expect to eliminate stars and low-z galaxies from our sample with an additional IR selection, we are able to use a color cut which is much broader (and closer to the stellar locus) than optical-only surveys such as SDSS. 

In addition to these color criteria, a candidate QSO is required to be unresolved in images taken through the two redder filters (eg. $g'$ and $r'$ for $z\sim3$ selection).  This minimizes contamination from low-z galaxies.   Our sample is limited to objects brighter than $r'< 22$, corresponding to a signal-to-noise ratio SNR $\geq 20$ in both $g'$ and $r'$.  Therefore, we have sufficient sensitivity to detect significant deviations from a point source (eg. $FWHM > 0.5''$).  

\subsection{Infrared Selection}
\label{ir_sel}
\begin{figure}
\epsscale{0.9}
\plotone{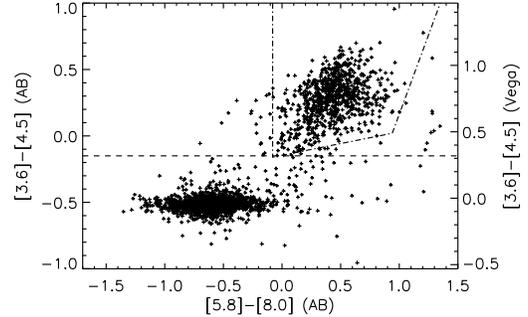}
\caption{IRAC colors of point sources in ELAIS-N1 with $r'<22$.  The dash-dotted line is the \citet{stern05} AGN selection and the dashed line is our selection $[3.6]-[4.5] > -0.15$ (AB) for demarcation of AGN from stars. \label{fig:pt_src_irac_color}}
\end{figure}

As seen in Figures \ref{fig:z3_opt_sel} and \ref{fig:z4_opt_sel}, we expect some contamination from stars in both our $U$-dropout and $g'$-dropout samples.  Fortunately, the mid IR SEDs of stars are very blue in color as they lie on the Rayleigh-Jeans side of the blackbody spectrum, whereas QSO SEDs are red, rising (in $\nu f_\nu$) towards longer wavelengths as seen in Figure \ref{fig:ir_template}.  Many groups have proven {\it Spitzer's} ability to select AGN with IRAC colors \citep{lacy04,stern05,hatziminaoglou05}.  In Figure \ref{fig:pt_src_irac_color}, we plot the IRAC colors of all point sources with $r' < 22$.  There are two clear loci of points.  The objects with blue IR colors (lower left) are stars and those with red IR colors (upper right) are AGN.  Both \citet{lacy04} and \citet{stern05} use all four IRAC bands to select AGN, but Figure \ref{fig:pt_src_irac_color} shows that the IRAC1-IRAC2 is a robust discriminator of AGN from stars when a point-source criterion is also used.  Also, we seek an IR selection using only IRAC channels 1 \& 2 since they are $\sim2$ magnitudes more sensitive than channels 3 \& 4 given the same exposure time (see Table \ref{tab:ir_data}).  Therefore, we apply the IR color cut of
\begin{equation}
[3.6]-[4.5] > -0.15 \  \  (AB)
\end{equation}
in addition to our optical selection to remove stars from our sample.  Here $[3.6]$ \& $[4.5]$ are AB magnitudes at 3.6 \& 4.5 $\mu$m respectively, corresponding to $f_{\nu}(4.5\mu$m$)/f_{\nu}(3.6\mu$m$) > 0.87$.

Finally, we search only sources with $f_{\nu}(4.5\mu$m$)\geq 10 \mu$Jy ($\sim 7\sigma$) so that Poisson errors in the flux measurement are minimized, thereby decreasing the risk of contamination from sources with large errors in IRAC colors.  We will see in Section \ref{ir_comp} that this matches well our $r' < 22$ optical cut.  

\begin{figure}
\epsscale{0.9}
\plotone{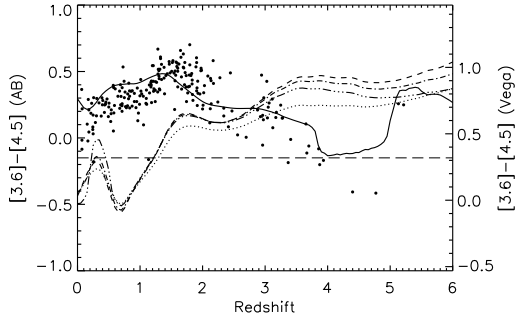}
\caption{$[3.6]-[4.5]$ color vs. redshift.  The line types are the same as in Figure \ref{fig:z3_opt_sel}.  Filled circles are SDSS QSOs in the SWIRE fields.  The lighter, long dashed line is our IRAC color selection, $[3.6]-[4.5] > -0.15 (AB)$. \label{fig:irac_color_z}}
\end{figure}

In addition to stellar contamination, some low redshift galaxies may  also meet our optical color criteria.  This is due to the Balmer- or 4000\AA-Break being mistaken for a $U$- or $g'$-dropout at $z<0.1$ and $0.1<z<0.5$, respectively.  Our point-source criterion will remove the lowest redshift galaxies as they will have large angular diameters, but our optical images may not resolve the most compact galaxies at $0.3<z<0.5$.  Therefore, we expect to see some contamination from galaxies within this redshift range in the $g'$-dropout sample.  In Figure \ref{fig:irac_color_z}, the $[3.6]-[4.5]$ color of our QSO and galaxy templates are plotted versus redshift, as well as the IR colors of SDSS QSOs in the SWIRE fields.  Unfortunately, many galaxies which may contaminate our $g'$-dropout optical selection (ie. $0.3<z<0.5$) are expected to meet our IR color selection. This is due primarily to the presence of strong 3.3 $\mu$m PAH emission in star-forming galaxies \citep{imanishi00}.  Therefore, we expect some contamination from $z\sim0.4$ galaxies within our $z\sim4$ QSO sample.  

In Figure \ref{fig:irac_color_z}, we can see that both our QSO template, and all but three of the SDSS QSOs remain above our IRAC color cut for $0<z<3.9$, including the 18 of 19 SDSS QSOs in the targeted redshift range for $U$-dropouts.  However, the QSO template dips close to our IR color cut between $3.8<z<4.8$, as do four of the five SDSS QSOs in this redshift range.  This is partly due to the change to bluer slope at $\lambda < 1.5\mu m$, but is caused primarily by H$\alpha$ redshifting into IRAC1.  At these redshifts the H$\alpha$ equivalent width is W(H$\alpha$)$\sim .15\mu m$ and the IRAC1 filter width is $0.7 \mu m$.  Therefore, the H$\alpha$ flux causes the $[3.6]-[4.5]$ color to decrease by $\sim0.2$ mags.  As a result, the $g'$-dropout selection may suffer from signifcant incompleteness between $3.9<z<4.5$ (in addition to the unreliability discussed above).

\subsection{Selection Summary}
\begin{itemize}
\item Point source brighter than $r'<22$ ($i'<21.5$ for $z\sim4$ sample) to ensure proper morphological characterization and bright enough to determine a large $U-g'$ ($g'-r'$) limit.
\item $IRAC2 > 10\mu$Jy to ensure high SNR needed for accurate mid-IR colors.
\item Optical colors which identify a Lyman Break in the continuum (see Table \ref{tab:opt_sel}).
\item Red mid-IR color ($[3.6]-[4.5]>-0.15$) to remove interlopers. 
\end{itemize}

\section{Results}
\label{results}
\begin{figure}
\epsscale{1.0}
\plotone{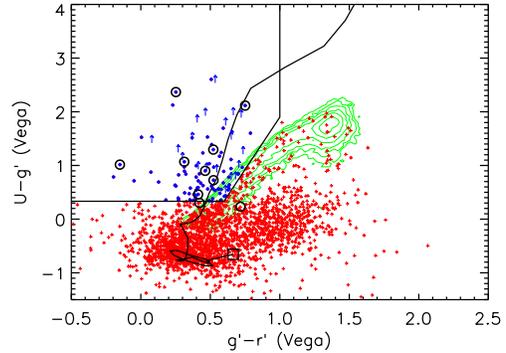}
\caption{The $U-g'$, $g'-r'$ optical colors of objects with $r' < 22$ and classified as point sources.  The solid curve is the color track of our QSO template with IGM absorption applied, and the square denotes the $z=0$ point.  The green contours denote the density (in color space) of objects categorized as stars by their blue IRAC colors ($[3.6]-[4.5]<-0.15$).  Red crosses are point sources with red IRAC colors ($[3.6]-[4.5]>-0.15$).  The blue arrows and circles also have red IRAC colors, but match the optical color criteria for $z\sim3$ QSOs.  The arrows denote upper limits where there is no detection in $U$.  Spectroscopically confirmed candidates are circled.  The photometry of two spectroscopically confirmed QSOs were revised and are therefore slightly out of our optical selection window but are still displayed here.  They are both QSOs at $z\sim3$ and their photometry is included at the end of Table \ref{tab:z3_cand}. \label{fig:z3_opt_col}}
\end{figure}

We performed our search in the EN1 and EN2 SWIRE fields, covering 11.7 deg$^2$ with both optical and IR coverage.  We found 100 $z\sim3$ and 26 $z\sim4$ QSO candidates which meet both our optical and IR criteria.  The $z\sim3$ candidates in EN1 and EN2 and their optical/infrared photometric data are listed in Table \ref{tab:z3_cand}.  The optical colors of the $z\sim3$ sample are plotted in Figure \ref{fig:z3_opt_col} along with the colors of all other point sources with $r'<22$.  We will show that the $z\sim4$ sample suffers from signifcant contamination and therefore do not list them.  In Section \ref{reliability}, we assess the reliability of our sample through spectroscopic follow-up.  In Section \ref{completeness}, we assess the completeness through Monte-Carlo simulations and derive effective volumes for use in computing the luminosity function.

\section{Reliability}
\label{reliability}
\subsubsection{Spectroscopic Follow-Up}
Optical spectra were obtained for 10 $z\sim3$ and 10 $z\sim4$ QSO candidates.  Thirteen spectra for faint candidates (6 $U$-dropouts and 7 $g'$-dropouts) were obtained with the Low-Resolution Imaging Spectrometer \citep[LRIS][]{oke95} on the Keck I Telescope during the nights of 03-04 March, 2005.  The $U$-dropout sample, with the expected redshift of $z\sim3$, have the most prominent emission lines at $\lambda<7500$ \AA.  Therefore, only the blue channel \citep[see Appendix in][]{steidel04} was used for these sources with a 1.5$''$ wide longslit and a 300 l/mm grism blazed at 5000 \AA, giving a resolution of 1.43 \AA\ pixel$^{-1}$ over $3300<\lambda<7650$ \AA.  For the $g'$-dropout sample, in order to detect the CIV (1549 \AA) line, we used a dichroic at 6800 \AA\ and used both the blue and red LRIS channel.  We used the same grism on the blue side and the 400 l/mm grating blazed at 8500 \AA\ on the red side giving a blue side resolution of 1.43 \AA\ pixel$^{-1}$ from $3300<\lambda<6800$ \AA\ and 1.86 \AA\ pixel$^{-1}$ from $7000<\lambda<8500$ \AA.  Total exposure times ranged from 5 to 15 minutes.  Both nights were photometric with seeing of $\sim 1.2''$.  The spectra were flux calibrated with observations of the standard star G1919B2B from \citet{massey88}.

Seven additional spectra (4 $z\sim3$ and 3 $z\sim4$ candidates) were obtained for brighter candidates with the COSMIC spectrograph on the 5-meter Hale Telescope at Palomar Observatory on the nights of 11-14 March, 2005.  A 300 l/mm grism blazed at 5500 \AA\ was used with a 1.5$''$ wide longslit, giving a dispersion of 3.1 \AA\ pixel$^{-1}$ and wavelength coverage of $3800<\lambda<8500$ \AA.  The nights were photometric with poor seeing ($\sim 2-4''$) so long exposure times of 10-60 minutes were required.  The spectra were flux calibrated with observations of the standard star G191B2B from \citet{massey88}.  

\begin{figure}
\epsscale{0.8}
\plotone{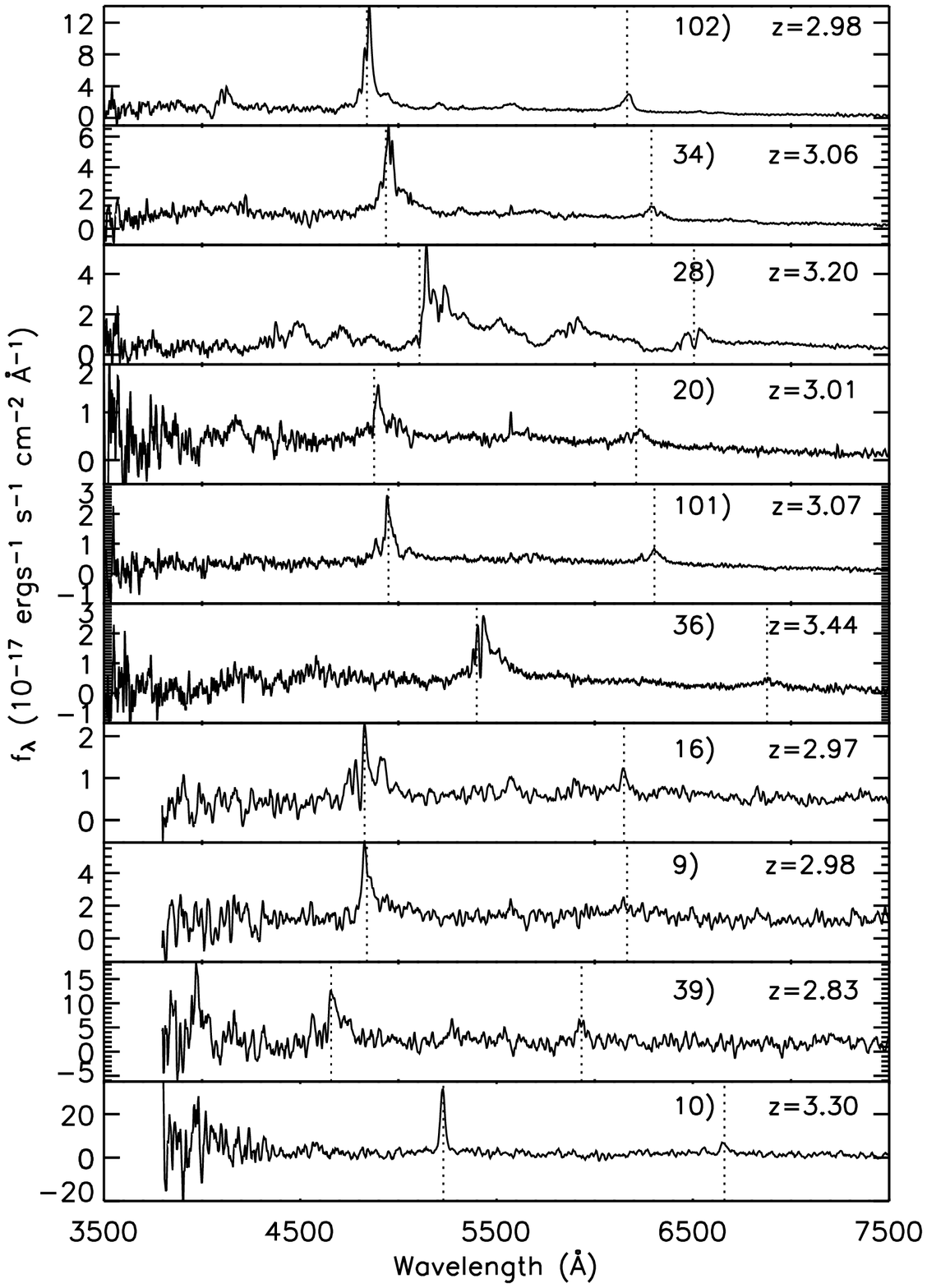}
\caption{Keck/LRIS (top 6) and Palomar/COSMIC (bottom 4) spectra of $z\sim3$ QSO candidates.  The ID numbers from Table \ref{tab:z3_cand} and redshifts are given and the Ly$\alpha$ and CIV lines are labeled. \label{fig:z3_spec}}
\end{figure}

All ten $U$-dropout candidates are QSOs with redshifts between $2.83<z<3.44$, the expected redshift range for our sample.  The spectra are shown in Figure \ref{fig:z3_spec}, exhibiting broad Ly$\alpha$ and CIV (1549\AA) lines and the QSOs with spectroscopic confirmation are circled in Figure \ref{fig:z3_opt_col}.  The six candidates observed with Keck/LRIS were chosen simply in ascending order in declination in order to sample the QSOs randomly in color space.  After these six were shown to be QSOs, we decided to target QSOs with extreme optical colors on the edges of our selection region as these are more likely to be interlopers or objects with spurious photometry.  In our subsequent run at Palomar, four of these QSOs with extreme colors were observed: one with blue $g'-r'$ color, one with red $U-g'$ color, and two with optical colors that lie close to stellar main sequence.  All four are in fact QSOs at the expected redshift.  

\begin{figure}
\epsscale{0.8}
\plotone{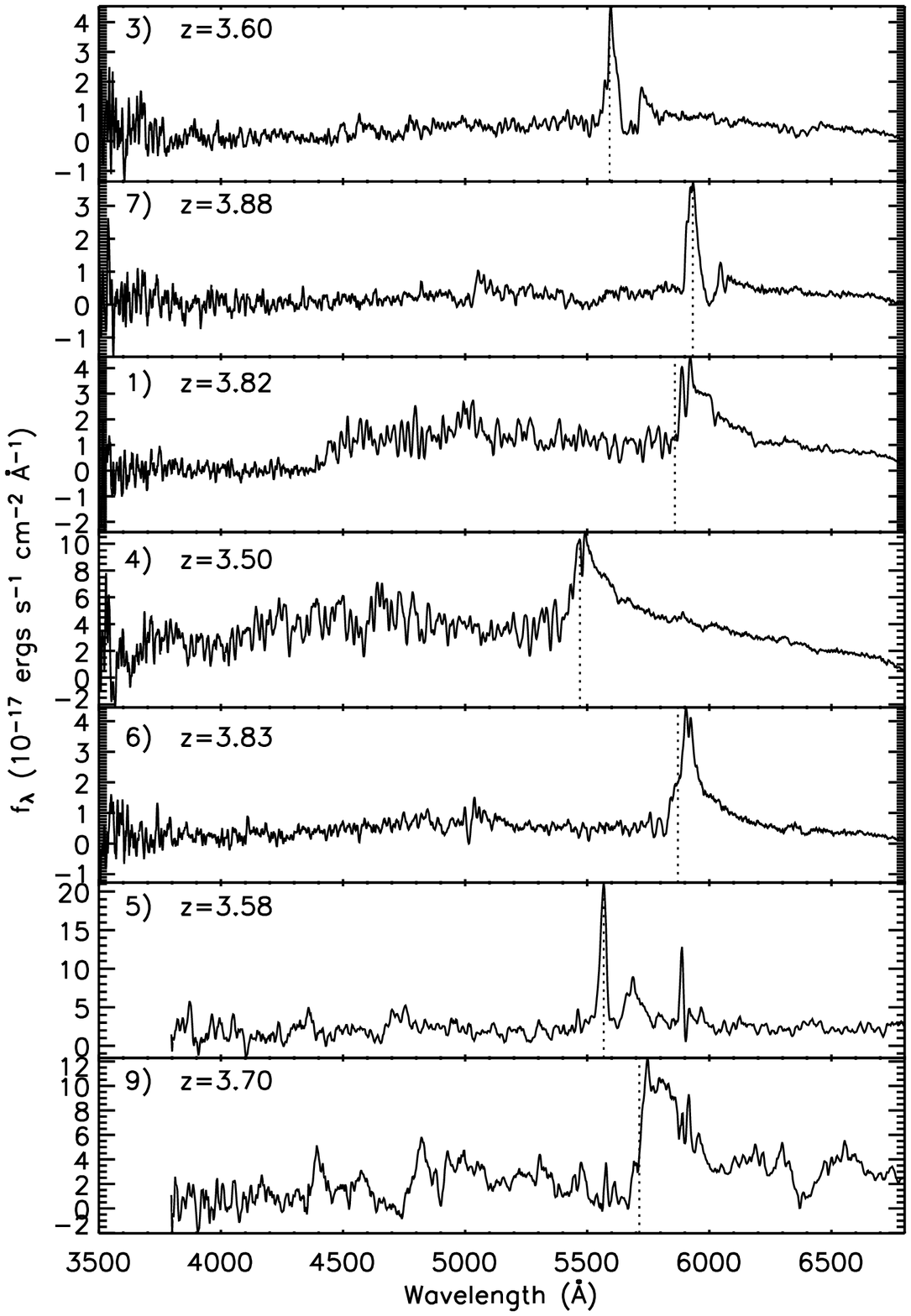}
\caption{Keck/LRIS (top 4) and Palomar/COSMIC (bottom 2) spectra of $z\sim4$ QSO candidates.  The ID numbers from Table \ref{tab:z4_spec} and redshifts are given and the Ly$\alpha$ line is labeled. \label{fig:z4_spec}}
\end{figure}

\begin{figure}
\epsscale{1.0}
\plotone{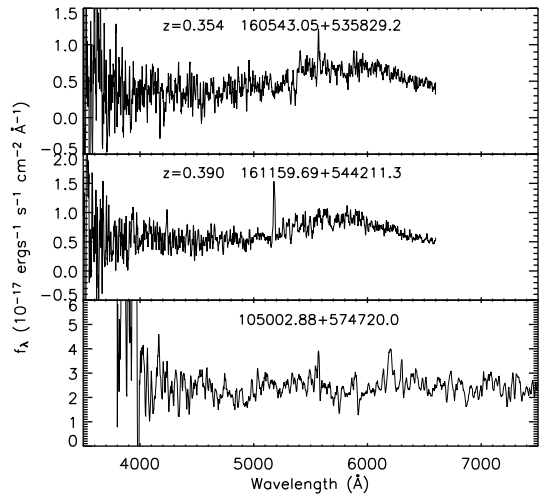}
\caption{Keck/LRIS (top 2) and Palomar/COSMIC (bottom) spectra of interlopers in the $g'$-dropout sample.  The official SWIRE names and redshifts (if known) are labeled. \label{fig:interloper_spec}}
\end{figure}

As expected (see Section \ref{ir_sel}), our $g'$-dropout sample suffers from significant contamination.  Seven of ten candidates are QSOs with $3.48<z<3.88$ and are displayed in Figure \ref{fig:z4_spec} and listed in Table \ref{tab:z4_spec}.  The three other spectra are plotted in Figure \ref{fig:interloper_spec}.  Two of the contaminants are galaxies at low redshift ($z=0.354$,$0.390$) and exhibit strong breaks in the continuum at $\lambda \sim 5500$ \AA.  At these redshifts, the interlopers' 4000 \AA\ break falls between the $g'$ and $r'$ filters.  The third contaminant has a high S/N spectrum, but did not show any emission features, nor a break in the continuum as expected, so we have not identified a redshift for this object.  Therefore, the reliability of our $z\sim3$ and $z\sim4$ QSO samples are 100\% ($>69$\% $1\sigma$) and 70$^{+16}_{-26}$\%, respectively.

The spectroscopic follow-up of the $g'$-dropouts shows the expected incompleteness at the high redshift end of the targetted range ($3.5<z<4.5$).  All 7 confirmed QSOs have $z<3.9$.  Beyond this redshift, H$\alpha$ redshifts into IRAC1 and may cause the $[3.6]-[4.5]$ color to be bluer than our color criterion.  

\subsubsection{Infrared Reliability}
\begin{figure}
\epsscale{1.0}
\plotone{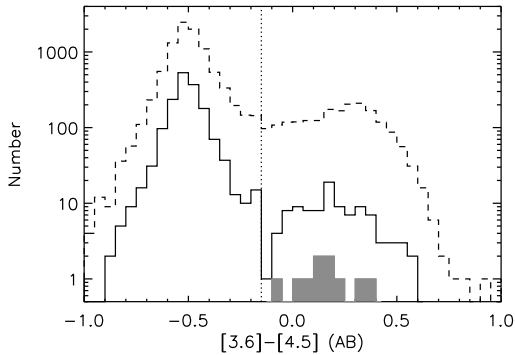}
\caption{The IRAC ($[3.6]-[4.5]$) color distribution of all point sources with $r'<22$ in our field (dashed line).  The solid line show the color distribution of point sources which also meet our optical color criteria.  The shaded histogram is the distribution of spectroscopically confirmed QSOs.  Our QSO sample is the portion of the solid histogram to the right of the vertical dotted line. \label{fig:irac_hist}}
\end{figure}

\begin{figure}
\epsscale{1.0}
\plotone{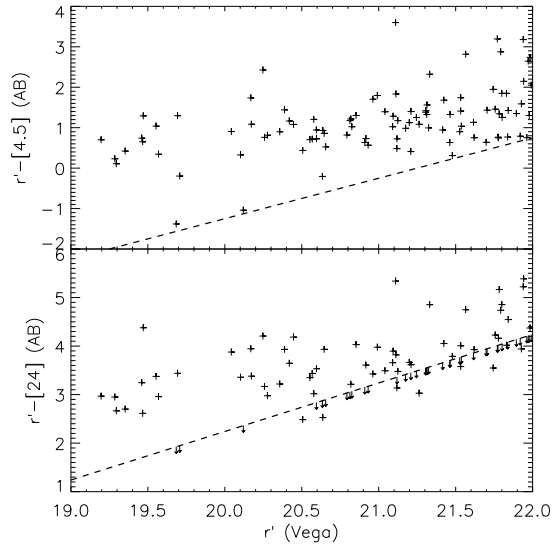}
\caption{Optical-IR color distribution of the $z\sim3$ QSO sample.  The x-axis is in Vega magnitudes since our magnitude bins are defined in Vega magnitudes and the y-axis is in AB magnitudes as it is simpler to interpret.  The dashed line in the top plot denotes the $f_{4.5\mu m}=10\mu$Jy limit of our search.  The dashed line in the bottom plot denotes the $f_{24\mu m}=250\mu$m completeness limit with arrows giving upper limits from non-detections. The four QSOs with $r'-[4.5] < 0.0$ are possible interlopers. \label{fig:opt_ir}}
\end{figure}

In Figure \ref{fig:z3_opt_col}, there are a few tens of point sources (out of more than 11,000) with red IRAC colors that lie within the stellar locus where we would not expect QSOs to lie (upper-right), suggesting possible contamination where the stellar locus crosses the color space of our QSO selection.  These may be red galaxies at moderate redshift, highly reddened QSOs, or stars whose IRAC colors are just redder than our color cut.  Figure \ref{fig:irac_hist} shows that indeed there is not a distinct bimodality of IRAC colors for point sources.  It's clear there are two peaks in the distribution, but there are hundreds of sources in the valley between the two peaks.  We have also plotted the IRAC color distribution of the point sources which also meet our optical color criteria.  In this histogram, nearly all of the sources in the valley between the two peaks have been removed, revealing a clear bimodality.  There are two reasons for the this.  Firstly, galaxies can have either red or blue IRAC colors depending on redshift, but should not lie in the optical color space of $z\sim3$ QSOs.  Secondly, though a few stars may have IRAC colors redder than our cut, the stars must be either very bright or very red to be detected in IRAC channel 2 at all.  Stars that lie in the $z\sim3$ optical color space have $r'-[4.5] = 1.6$ mags (AB).  Since the IRAC channel 2 flux limit corresponds to 21.4 in AB magnitudes, this means we shouldn't detect the stars in channel 2 unless they are brighter than $r' \lesssim 19.7$ (Vega).  Indeed, in Figure \ref{fig:opt_ir} there are two sources at the bright end of our sample that have blue optical-IR colors and may be stars.  

\begin{figure}
\epsscale{1.0}
\plotone{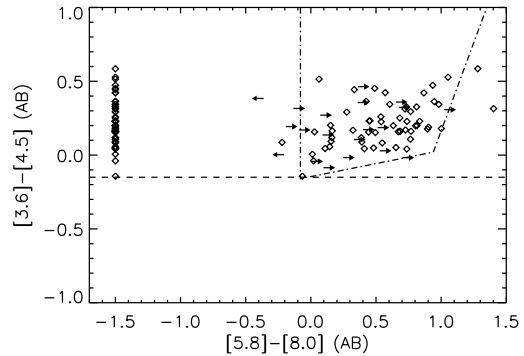}
\caption{IRAC colors of the $z\sim3$ QSO sample.  The lines are as in Figure \ref{fig:pt_src_irac_color} .  Red arrows are lower limits to the $[5.8]-[8.0]$ color based on non-detections in IRAC3.  Blue arrows are upper limits to the $5.8]-[8.0]$ color based on non-detections in IRAC4. Non-detections in both IRAC3 \& IRAC4 are plotted along the left hand side.\label{fig:qso3_ir_color}}
\end{figure}

As mentioned in Section \ref{ir_sel}, IRAC color selection of AGN is robust when all four bands are used because the SEDs of most contaminants are not flat (or rising toward longer wavelengths) over such a broad wavelength range.  Therefore, we can use all four IRAC fluxes, when available, to assess the reliability of our single color AGN selection.  In our sample, 58\% (58 of 100) of our objects are detected in all four IRAC channels and their IRAC colors are plotted in Figure \ref{fig:qso3_ir_color}.  Essentially all of these objects have IR colors in the expected locus of AGN.  Therefore, it appears the contamination rate is low.   However, as stellar contaminants are expected to have blue IRAC colors, we do not expect to detect the fainter interlopers in IRAC3 and IRAC4 and we cannot assess the nature of these objects.  Therefore, we look at only the bright end of our sample where all sources are detected in all four bands.  At $r' < 20.5$, 21 of 24 (88\%) sources are detected in all four IRAC bands, and all of these have IR colors within the expected AGN locus.  If we make the conservative assumption that the three non-detections (in IRAC3 \& IRAC4) amongst the bright sample are not QSOs, then we obtain an upper limit to the contamination rate of $\lesssim$13\%.  

In summary, the  point sources which match our optical color criteria have a distinct bimodal distribution in $[3.6]-[4.5]$ colors, suggesting minimal contamination from stars, except at the bright end ($r' < 20$).  However, the four-band IRAC colors suggest that even the bright end of our sample has a contamination rate of less than 13\%.  

\section{Completeness}
\label{completeness}

\subsection{Infrared Completeness}
\label{ir_comp}
As mentioned in Section \ref{ir_sel}, we expect our IR color cut to include nearly all QSOs at $z\sim3$ (as seen with the SDSS QSOs).  However, we must ensure that no QSOs are missing due to the $f_{\nu}(4.5\mu$m$) \geq 10 \mu$Jy.  Figure \ref{fig:opt_ir} plots the optical-IR colors of all $U$-dropout candidates as a function of the optical magnitudes.  The locus of colors agree well with the infrared selected QSOs of \citet{brown06}.  Also plotted is the limit due to the IRAC2 flux cut.  Except for the faintest of the six half-magnitude bins, there is is no incompleteness due to non-detections in IRAC2.  In the faintest bin, $21.5<r'<22.0$, it is possible that we miss a few of the bluest $r'-[4.5]$ QSOs, but we estimate this to be $<15$\% (based on the $r'$-[4.5] distribution in Figure \ref{fig:opt_ir}) and make no attempt to correct for it.  We've also verified that increasing the counts by 15\% in this bin has a neglible effect on the QLF fit.

\subsection{Morphological Completeness}
\label{morph_comp}
Our selection requires that the candidates be unresolved in the optical data, where typical resolution is $\sim 1.1''$.  We have assumed that the QSOs are significantly brighter than their host galaxies and, even if we could detect the host galaxy, galaxies at $z>3$ should have angular diameters less than 1$''$ \citep{giavalisco02}.  To test this assumption, we've matched our catalogs to all of the SDSS QSOs (which do not have to meet a morphological criterion) within the EN1 and EN2 fields.  Of the 58 SDSS QSOs at high redshift ($z>1$), 57 (98\%) are categorized as point sources in our optical catalogs.  Therefore, we do not expect any significant incompleteness due to our point-source criterion.  

\subsection{Optical Completeness}
Our color selection should be strict enough to minimize contamination, but broad enough to encompass the majority of the targeted QSOs to minimize completeness corrections.  As a first-order look at completeness, we compare with the SDSS QSO sample between $3.1<z<3.2$ where SDSS is $\sim 80$\% completete \citep{richards06}.  SDSS uses four colors, to look at all point-sources away from the color space of stars \citep{richards02}.  Therefore it is possible that SDSS selects redder QSOs at $z\sim3$, which would lie outside of our color selection.  Converting SDSS to Vega magnitudes, and correcting for slight $U$/$u'$ filter differences (0.06 mag correction), $\sim$97.5\% (392/402) of SDSS QSOs at $3.1<z<3.2$ would also be selected with our color selection.  Therefore, $<3$\% of SDSS QSOs are redder, and selected with other colors in the SDSS filters.  

Significant dispersion in spectral features may cause some QSOs to lie outside of our selection criteria.  The most important factors affecting the optical color (and therefore the completeness) are redshift, intervening high column-density HI absorbers, UV continuum slope, and emission line equivalent widths, and variability.  Here we present our Monte Carlo simulations to assess the combined effects of these characteristics on our completeness as a function of magnitude and redshift.  We then use this completeness to derive the effective volume of our sample as a function of apparent magnitude. 

\subsubsection{Model of QSO Optical Color Distribution}
\label{sim}
Both the UV spectral slope of the continuum $\alpha_{\nu}$, where $f_\nu \propto \nu^\alpha$, and the emission line equivalent widths, $W$, can vary about the mean and cause significant dispersion in colors at any given redshift.  The mean of the spectral slope and the equivalent widths are defined by our template spectrum and are consistent with values found by \citet{vanden_berk01} and \citet{hunt04}.  The distribution of these attributes are assumed to be gaussian with standard deviations taken from \citet{francis96}, and are listed in Table \ref{tab:sim_params}.  

Along the line-of-sight to any given QSO at high redshift there are hundreds of intervening neutral hydrogen clouds (or filaments) absorbing the UV light through Lyman line and Lyman continuum absorption.  Much of this absorption is caused by the rare, high column-density absorbers known as Lyman Limit Systems (LLSs, N$_{HI} > 1.6\times 10^{17}$cm$^2$) and Damped Lyman-$\alpha$ Systems (DLAs; N$_{HI} > 1\times10^{20}$ cm$^2$).  Therefore, there is a large dispersion in line-of-sight HI opacity at any given redshift due to the small numbers of these high column density clouds.  \citet{bershady99} show that simple analytic expressions for the mean and scatter in line-of-sight opacity are insufficient and, therefore, Monte-Carlo simulations are required to correctly represent the stochastic distributions of the absorbers.  

The number density distribution of HI absorbers is typically given as a fit to a power-law with redshift

\begin{equation}
\label{eqn:nz}
N(z) = N_0(1+z)^\gamma
\end{equation}

{\noindent}and the column density distribution is given as

\begin{equation}
\label{eqn:nh}
\frac{df}{dN_{HI}}\propto N^{-\kappa}_{HI}
\end{equation}

{\noindent}Since absorbers of different column densities are known to evolve differently \citep{kim97}, we have split them into four groups with differing evolution, similar to the work of \citet{bershady99} and \citet{fan01} but with updated values for DLA number densities.  The parameters used are summarized in Table \ref{tab:laf_params}.  

For each line of sight, the number of line-of-sight absorbers is selected from a Poisson distribution with the expectation value set to $<N>$ of the population.  

\begin{equation}
<N> = \int_0^{z_{QSO}} N(z)dz
\end{equation}

\begin{figure}
\epsscale{1.0}
\plotone{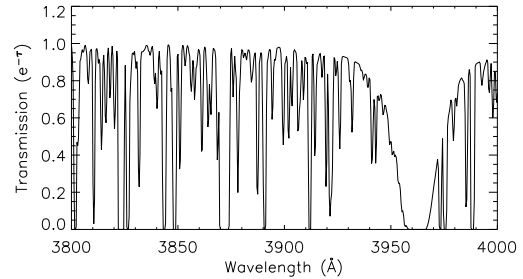}
\caption{A subsection of the simulated IGM transmission ($e^{-\tau}$) curve for one simulated line-of sight.  All aborption features are fully resolved in the simulation. \label{fig:igm_z3_ex}}
\end{figure}

{\noindent}Their redshifts and column densities are chosen randomly from the distributions in Eqns. \ref{eqn:nz} and \ref{eqn:nh}, respectively.  Finally, we compute Voigt profiles with natural broadening and doppler widths, $b$, from Table \ref{tab:laf_params} for the first ten Lyman lines for each absorber.  We choose not to model the distribution of Doppler widths (or its evolution) as this is expected to be a small, second order effect.  Continuum absorption is applied with a scattering cross-section of $\sigma_0 = 6.3\times10^{-18}$ cm$^2$ at the Lyman Limit and decreasing as $\nu ^{-3}$.  An example of the Ly$\alpha$ Forest transmission for one line of site at $z=3$ is plotted in Figure \ref{fig:igm_z3_ex}.   

For 300 lines-of-sight in each redshift bin of $\Delta z = 0.1$, we compute QSO spectra with spectral slopes and emission line equivalent widths culled from the distributions defined in Table \ref{tab:laf_params} and apply the simulated IGM absorption for that line of sight.   

QSOs of these luminosities and redshifts exhibit variability of order $\Delta m \sim 0.15$ mags \citep{vanden_berk04}.  29\% of our fields were observed at separate times (over two years) in the $U$ and $g'$ filters, and 29\% of our fields were observed at separate times in the $g'$ and $r'$ filters.  We therefore add a variable offset selected from a gaussian distribution with $\Delta m \sim 0.15$ to the computed $U-g'$ and $g'-r'$ optical colors of 30\% of the simulated lines of sight.  The net effect of this variability is a slight $\sim 5$\% decrease in completeness, and a slightly broader redshift range.  

After adding in variability and photometric errors, we compute the optical colors for the 300 QSOs in each redshift bin.  This allows us to determine the selection completeness based on color selection alone, but the imaging depth must still be taken into consideration.  

The optical magnitude limit of our QSO search, $r'=22.0$, is set by our infrared depths and corresponds to $\sim 20\sigma$ detections in both $g'$ and $r'$.  Therefore, we do not expect any incompleteness due to non-detections in $g'$ or $r'$.  However, when looking for significant ``$U$-dropouts'', it's important to have deeper imaging in $U$ relative to the bands at longer wavelengths.  Otherwise, non-detections will result in upper limits in flux which cannot distinguish between a true ``dropout'' or a red source which is just below the detection limit.  The WFS was not designed for finding $U$-dropouts and the $U$ band is generally less sensitive due to CCD quantum efficiency (QE) and poor throughput in the telescope optics.  Therefore, the $U$-band images in the SWIRE fields are less sensitive (to all classes of object) than the $g'$ and $r'$ images.  Furthermore, this sensitivity varies by $\sim 0.5$ magnitudes from pointing to pointing due to changing observing conditions (seeing, airmass, lunar phase).  Therefore, careful measurements of the selection completeness as a function of pointing must be assessed, along with dependencies on redshift and magnitude.  

Both EN1 and EN2 are comprised of 54 22.8$'$$\times$11.4$'$ optical pointings.  For each CCD at each pointing, a $5\sigma$ limiting magnitude is determined within our $2.3''$ diameter aperture and is used as an upper limit when no object is detected.  

\begin{figure}
\epsscale{1.0}
\plotone{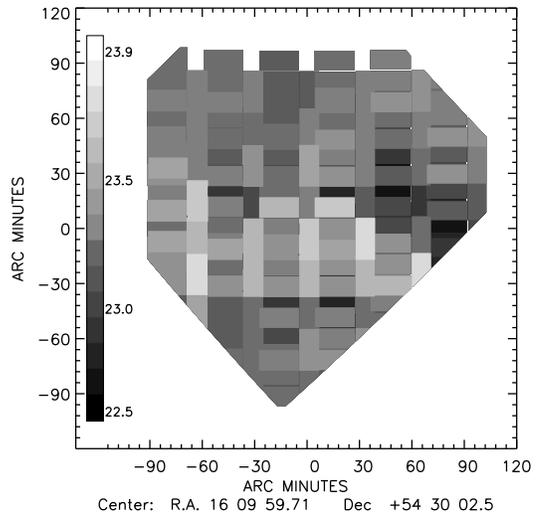}
\caption{The U-band depth map for EN1.  The grey-scale represents $U$ magnitude limits in intervals of 0.1 mags from 22.5 to 24.0 (Vega) with most of the area between 23.0-23.5 \label{fig:en1_u_depth}}
\end{figure}

Due to overlaps in pointings and varying observing conditions, it is nontrivial to compute the limiting magnitude for a given position on the sky.  For both fields we make ``depth maps'' by making mosaic images of the entire field with pixel values set to the $5\sigma$ limiting magnitude of the deepest pointing which covers that pixel.  In this way, we can quickly compute the area over which we can detect an object of a given magnitude $U_{test}$ by summing the pixels with value greater than $U_{test}$.  These depth maps are trimmed to the area which also has {\it Spitzer} data.  The U-band depth map for EN1 is plotted in Figure \ref{fig:en1_u_depth}, showing the non-uniformity and complex structure of the coverage.  

As shown in Figure \ref{fig:z3_opt_sel}, the bluer of the two optical colors, ($U-g'$), becomes redder as the QSO goes to higher redshift, making non-detections in $U$ increasingly likely at higher redshift.  For example, a typical $r'=21$ QSO at $z=2.9$ will have $g'\sim21.5$ and $U\sim 21.8$, bright enough to be detected in all bands by our survey.  However, a QSO with the same $r'$ magnitude at $z=3.4$ will have $g'\sim 22$ and $U \sim 24.5$, too faint to be detected in our $U$ images.  Our upper limit in this case is insufficient in distinguishing a high-redshift QSO candidate from a low-redshift object with bluer $U-g'$.  

Given our simulated spectra and our depth maps, we determine an effective completeness in the following manner.  For each redshift bin, we compute $U-g'$ and $g'-r'$ colors for the 100 simulated QSOs along 100 different lines of sight.  Then we compute the percentage of these QSOs which would be selected by our color criteria if our imaging was sufficiently deep.  This gives us a measure of our completeness based strictly on our color criteria alone, $C_{color}$.  The completeness cutoff at the high redshift end ($z\sim3.5$) is nearly a step-function since the QSO color track moves perpendicular to our color cuts and most of the dispersion in color is parallel to the color cut.  At the low redshift end ($z\sim 2.9$), the incompleteness is predominantly due to line-of-sight variations in IGM.

\begin{figure}
\epsscale{1.0}
\plotone{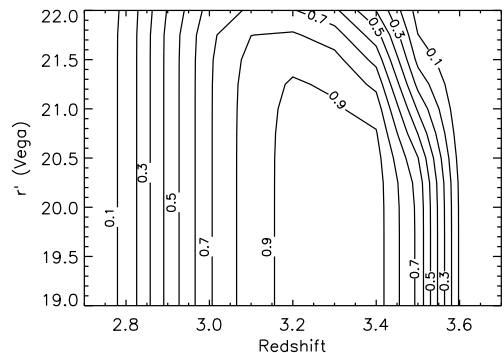}
\caption{The $z\sim3$ QSO completeness $C(r',z)$ contours as a function of apparent $r'$ magnitude and redshift.  The contours are spaced at $\Delta C = 0.1$ intervals. \label{fig:opt_comp}}
\end{figure}

Of the QSOs which meet the color criteria, the fluxes are scaled to give the desired $r'$ magnitude in intervals of $\Delta r' = 0.25$ mags and the $g'-r'$ color is used to determine the $U$-band depth required to either detect this QSO or derive a lower limit to the magnitude which is high enough to put it in the color-color selection window.  The percentage of QSOs with colors in our selection window which would also be selected given the $U$-band depth at that pixel value is denoted as $C(r',z)$, plotted in Figure \ref{fig:opt_comp}.  The effective volume of the survey can then be calculated as 

\begin{equation}
V_{eff}(r') = d\Omega \int_{z=0}^{z=\infty}C(r',z)\frac{dV}{dz}dz
\end{equation}

{\noindent}where $\Omega$ is the solid angle of the survey and $dV/dz$ is the differential comoving volume.  The effective volumes and average redshifts for each half-magnitude bin are given in Table \ref{tab:tab_qlf}.  To give an idea of the scales of the incompleteness corrections, the effective volume in our faintest magnitude bin $V_{eff}(r'=21.75)$ is 74\% of the effective volume in our brightest bin $V_{eff}(r'=19.25)$, requiring a relatively small correction of 35\% to the number counts.

\section{The $z\sim3$ QSO Luminosity Function}
\label{qlf}

Given the effective volume, $V_{eff}(r')$, the comoving space density is then

\begin{equation}
\Phi(r')=\frac{N(r')}{V_{eff}(r')}\frac{1}{w_{bin}}.
\end{equation}

{\noindent}where $N(r')$ is the number of QSOs in the $r'$ bin and $w_{bin}$ is the width of the bin in magnitudes.  We then convert $r'$ to the absolute AB magnitude at 1450\AA\ since the $r'$ filter covers 1450 \AA\ over our entire redshift range and this value is generally used for high redshift QSO studies.

\begin{equation}
M_{1450} = r' + r'_{AB}(Vega) - DM(z=3.2) + 2.5log(1+3.2) + K(\Delta z)
\end{equation}

{\noindent}where $r'$ is the Vega magnitude listed in Table \ref{tab:z3_cand}, $r'_{AB}(Vega)=0.15$ is the Vega to AB conversion, $DM=47.19$ is the distance modulus at z=3.2, and $K(\Delta z)$ is the K-correction resulting from shifts in redshift around $z=3.2$.  As we don't have exact redshifts for most of our QSOs, we set $K(\Delta z)=0$, but note that this varies by $\pm0.1$ mags from $2.9<z<3.5$.  

\begin{figure}
\epsscale{1.0}
\plotone{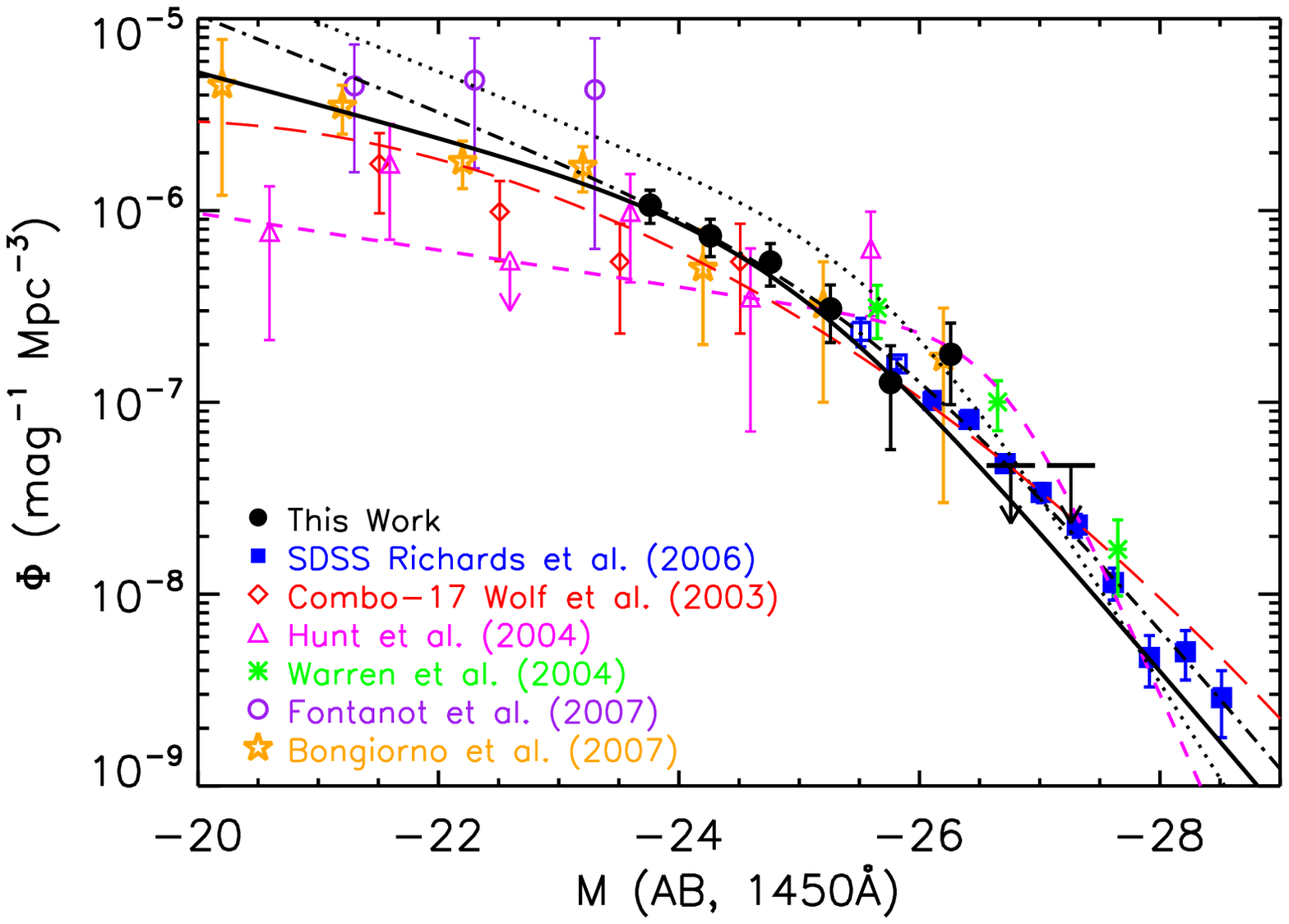}
\caption{The $z\sim3.2$ QSO luminosity function. The binned data are plotted from this work (black circles and black arrows for upper limits), SDSS \citet[][blue squares]{richards06}, COMBO-17 \citep[][red diamonds]{wolf03}, \citet[][cyan triangles]{hunt04},WHO \citep[][green asterisks]{warren94}, \citet[][open purple circles]{fontanot07}, and \citet[][orange stars]{bongiorno07}.  Filled symbols were used in the fitting of the QLF.  Also plotted are the fitted QLFs using the binned data (dot-dashed line), the maximum likelihood fit with Pure Luminosity Evolution (solid line), and the QLFs of \citet[][dotted line]{pei95}, \citet[][magenta dashed line]{hunt04}, and \citet[][red long dashed line]{wolf03}. \label{fig:qlf}}
\end{figure}

The resulting QSO luminosity function at $z\sim3.2$ is plotted in Figure \ref{fig:qlf}, along with previous surveys of QSOs at these redshifts.  Our measurements, while using smaller magnitude bins ($w_{bin} = 0.5$ mags) than previous surveys of faint high-z QSOs, have significantly reduced error bars.  The space densities at the bright end of our survey match well with the faintest bins from the SDSS \citep{richards06} and show a clear transition to a shallower slope at the faint end.  We use our data in combination with the \citet{richards06} SDSS results at $z\sim3.25$  because it is the largest sample available, and its completeness corrections have been carefully determined.  

We do a least-squares fit to the standard double power-law given in Eqn. \ref{eqn:qlf_lum}, converted to absolute magnitudes

\begin{equation}
\Phi(M_{1450},z) = \frac{0.92 \times \Phi(M_{1450}^*)}{10^{0.4(\alpha+1)(M_{1450}-M_{1450}^*)}+10^{0.4(\beta+1)(M_{1450}-M_{1450}^*)}}.
\label{eqn:qlf_mag}
\end{equation}

{\noindent}As mentioned previously, when the error bars are reduced, the quasar luminosity function exhibits curvature over all luminosities, not just at the break \citep{wolf03,richards05}.  Therefore, since both our data and the SDSS data show some curvature, and the SDSS has much smaller error bars, a least squares fit of a double power-law will force the break position ($M^*$) to be contained within the range of magnitudes covered by the SDSS.  This produces a very steep faint end slope and is a poor fit to our faintest bins.  Of the four parameters that determine the QLF, the SDSS can only measure the bright-end slope with certainty at $z\sim3.2$.  Therefore, we also perform a fit with a fixed bright end slope, $\alpha = -2.85$, defined by the SDSS measured relation of $\alpha$ with redshift at $z=3.2$.  This yields a more realistic values of $M^*=-25.6$, and a faint end slope, $\beta = -1.62\pm0.19$, which is constrained by 4-5 bins fainter than $M^*$.  The results of the double power-law fits are given in Table \ref{tab:qlf_params}. 

\subsection{Maximum Likelihood Fit}

In the above fit to the binned data, we assumed that all of our QSOs are at a single redshift ($z=3.2$).  For most of our objects, however, we do not have spectroscopic redshifts and do not know the absolute magnitudes.  This can be problematic because we do not know, for example, if the brighter QSOs are indeed more luminous or simply at the low redshift end of our redshift range (the distance modulus changes by $\sim$ 0.5 mags from $z=2.8-3.4$).  Furthermore, the luminosity function is known to change over these redshifs, which may skew the apparent magnitude distribution from what is expected at a single redshift.  

Our only known quantity is the distribution of {\it apparent} magnitudes of QSOs in this redshift range.  Therefore, we have modeled the expected distribution in QSO apparent magnitudes to compare to the observed distribution.  The model distributions are computed in the following way.  

\begin{enumerate}

\item We allow $\Phi^*$, $M^*$, and faint-end slope, $\beta$, to vary independent of one another.  We keep the bright-end slope fixed at $\alpha = -2.85$ as determined by SDSS QSOs \citep{richards06}.

\item For each set of parameters, we compute the apparent magnitude distribution in small redshift intervals. 

\item At each redshift, we apply the completeness function as a function of apparent magnitude, $C(r',z)$, computed in Section \ref{sim}.

\item Finally, we sum up the apparent magnitude distribution in each redshift interval to determine the expected apparent magnitude distribution function over the entire redshift range for each set of QLF parameters. 

\item We repeat the above steps three times with different QLF evolution: no evolution, pure luminosity evolution (PLE) and pure density evolution (PDE).  For the two evolving models, we choose the level of evolution to fit the variation in space density of bright QSOs seen by SDSS \citep[$\sim$ 40\% decrease from $z \sim 2.8 - 3.5$,][]{richards06}.

\end{enumerate}

In addition to the observed apparent magnitude distribution from our sample, we also use the QSO sample of \citet{richards06} from the SDSS Data Release 3.  We only use the QSOs with $2.9 < z < 3.5$ and use the completeness for each QSO derived in \citet{richards06}.  

The apparent magnitude distribution function gives the relative probability of finding a QSO with a given magnitude from the whole sample.  We compute the likelihood that a given parameter set adequately describes the data as the product of the values of the apparent magnitude distribution function for all of the QSO $r'$ magnitudes in our sample \citep{marshall83}.  We then find the set of parameters which maximizes this likelihood.  The best fit model parameters are given in Table \ref{tab:qlf_params}.  The maximum-likelihood fit gives a somewhat shallower faint-end slope, $\beta = -1.42 \pm 0.15$, and a similar location of the break, and agrees within $\sim 1\sigma$ with the binned QLF when we assumed all QSOs were at $z=3.2$. The apparent magnitude distributions which included evolution in the QLF (both PLE and PDE) produce best-fit parameters which were not significantly different from the no evolution fits because there is little evolution over such a small redshift range.  We choose to use the PLE model, with $\beta = -1.42 \pm 0.15$, in the subsequent analysis.

\section{Comparison with Previous Luminosity Functions}
\label{qlf_comp}
Plotted in Figure \ref{fig:qlf} are three previously determined luminosity functions at $z\sim3$.  \citet{pei95} compiled several QSO samples to produce a QLF and its evolution between $0<z<4.5$.  A constant QLF shape was assumed, and fitted with Pure Luminosity Evolution.  Unfortunately, most of the high redshift QSOs are very luminous, so the faint-end slope is mostly determined by lower redshift QSOs.  As seen in Figure \ref{fig:qlf}, the \citet{pei95} QLF has a steeper faint end slope, $\beta = -1.6$ and lies above our QLF determination until $M^*_{1450} < -26$, where it passes through the SDSS data points.  We believe that this discrepancy is caused by the assumption of a constant QLF shape.  The \citet{pei95} QLF shape was determined mostly by QSOs at $z<3$, and normalized to the bright QSOs at high-redshift.  Because the bright end slope at $z > 3$ is getting shallower \citep{richards06}, this has caused a significant over-estimate of the space densities of faint QSOs.  In fact, all previous QLF estimates at high redshift that assume a shape derived at low redshift and normalize using bright, QSOs will overestimate the number of faint QSOs at $z>3$.  At $z\sim3$, this results in a factor of $\sim2$ overestimate, but will get worse at higher redshift as the bright end slope is measured to get even shallower at $3<z<5$.

The H04 QLF, though measured with a much smaller sample, shows a very shallow faint-end slope and a steep bright-end slope, resulting in a very prominent break.  Our QLF has a significantly shallower bright-end slope and a steeper faint end slope.  The space densities of our faintest QSOs are about twice that of H04 and predict 2-3 times more QSOs between $-24<M_{1450}<20$.  

H04 threw out most of their AGN sample (16 of 29) because the emission line widths were less than 2000 km/s. These narrow-line AGN are, on average, one magnitude fainter than the broad-line sample.  Therefore, if these were included, they would have added significantly to the faint-end counts, resulting in a steeper faint-end slope.  We cannot discriminate between different types of AGN in our sample,  since we only have spectra for 10 objects (9 of 10 have $FWHM > 2000$ km/s).  However,  narrow-line AGN (ie., $FWHM < 2000 $ km s$^{-1}$) may explain at least part of the discrepancy in our faint-end slopes.  

Recent surveys of X-ray selected AGN have also concluded that the faint end slope gets shallower at higher redshift \citep{ueda03,hasinger05}.  The \citet{hasinger05} sample overlaps our redshift range but differs from ours in the same way as H04, as it is a soft X-ray selected type-I AGN sample.  If the difference between the H04 and \citet{hasinger05} luminosity functions and our QLF is attributed to increasing numbers of moderately obscurred AGN at fainter rest-frame UV luminosities (as suggested by \citet{ueda03}), then we would expect to see increasing numbers of QSOs with redder UV spectral slopes and higher $L_{IR}/L_{UV}$ ratios amongst faint QSOs.  Indeed, Figure \ref{fig:opt_ir} shows a population of $r'>21$ QSOs which are redder, in both $r'-[4.5]$ and $r'-[24]$, than any at $r'<21$, which may not be included in the H04 or \citet{hasinger05} LFs. 

A recent optical/X-ray search for faint AGN at $3.1<z<5.2$ by \citet{fontanot07} finds relatively large space densities, requiring a steep faint-end slope, $\beta = -1.71 \pm 0.41$.  Though the errors are large, this rules out a significantly shallower faint-end slope at $z>3$ and agrees, within $1\sigma$, with our fit.  In their study, only 18\% (2/11) AGN have narrow lines.  Therefore, the fraction of narrow-line to broad-line AGN is too small to completely explain the discrepancy between the shallow faint-end slope of H04 and the steep faint-end slope found in this study and in \citet{fontanot07}. 

The COMBO-17 \citep{wolf03} and VVDS \citep{bongiorno07} QLFs \citet{bongiorno07} are both consistent with our QLF, within errors.  Both of these surveys did not require any optical color or morphological criteria yet they still agree with our numbers.  Both the \citep{wolf03} and \citep{bongiorno07} samples are larger than ours (192 and 130, respectively), but the QSOs are spread out over all redshifts ($0<z<5$) so they have 5-10 times fewer QSOs in this redshift range.  

\section{QSO Contribution to HI Ionizing Flux at $z\sim3.2$}
\label{ion_bkg}

With this new determination of the QSO Luminosity Function, we can simply integrate $\Phi(L)$ given in Equation \ref{eqn:qlf_lum} to determine the specific luminosity density (at $\lambda=1450$\AA) of QSOs at $z\sim3.2$,

\begin{equation}
\epsilon = \int \Phi(L)\ L\ dL.
\label{eqn:eps_uv}
\end{equation}

Integrating from $-30<M_{1450}<-20$, ($43.96<$ log$(L_{1450})<47.96$ ergs s$^{-1}$), we derive a value for the specific luminosity density, $\epsilon_{1450} = 7.3\times10^{24}$ ergs s$^{-1}$ Hz$^{-1}$ $h$ Mpc$^{-3}$.  This is comparable to the values derived from the \citet{hunt04} QLF ($7.1\times10^{24}$) and significantly lower than the value from the \citet{pei95} QLF ($1.4\times10^{25}$) when correcting for different cosmologies.  Although our QLF determination predicts more integrated UV flux from faint QSOs than that of \citet{hunt04}, our $M^*$ is more than 1.5 magnitudes fainter.  These two effects essentially cancel out to produce similar luminosity densities.

We can now determine the photoionization rate

\begin{equation}
\label{eqn:gamma}
\Gamma = \int_{\nu_0}^{\infty}d\nu 4\pi \frac{J(\nu)}{h\nu}\sigma_{HI}(\nu).
\end{equation}

{\noindent}However, we can not assume that $J(\nu)$ is the shape of the average QSO SED, since the higher energy photons will be reprocessed by HI and HeII, resulting in a higher value for $\Gamma$.  \citet{haardt96} have modeled this reprocessing in a ``clumpy'' IGM to determine the effect on the HI photoionization by QSOs.  This correctly includes HI clouds as sources of ionizing photons, as well as sinks, and increases $\Gamma$ by $\sim40$\% at $z\sim3$.  We multiply $\epsilon_{1450}$ by the ratio of $f_{912}/f_{1450}=0.58$ in our template, and convert to a {\it proper} volume emissivity at $z=3.2$ to directly compare with the $\epsilon_{Q}$ calculated by \citet{haardt96} and how it scales with the photoionization rate, $\Gamma_{HI}$ and the ionizing intensity at the Lyman Limit $J_{912}$.  It should be stated that the scaling relations from $\epsilon_Q$ to $J_{912}$ and $\Gamma_{HI}$ are dependent upon the value of $\epsilon_Q$ itself, as this will effect the ionization levels of the surrounding medium.  However this is a secondary effect, and our $\epsilon_Q$ is within $\sim50$\% of the \citet{haardt96} value so we don't expect this to be a large effect on the scaling relations.  

We get values of $\Gamma_{HI} \sim 4.5\times10^{-13}$ s$^{-1}$ and $J_{912} = 1.5\times 10^{-22}$ ergs s$^{-1}$ cm$^{-2}$ Hz$^{-1}$ sr$^{-1}$.  As pointed out by \citet{hunt04}, these should be taken as upper limits as it assumes that all ionizing photons escape from QSOs of all luminosities, though this is not at all clear for lower luminosity AGN. 

It is interesting to compare the HI photoionization rate from QSOs with that from star-forming galaxies (LBGs at these redshifts).  Unfortunately, it is difficult to determine the photoionization rate from Lyman Break galaxies, since their photoionizing SEDs are difficult to directly detect and are sensitive to parameters with large uncertainties: dust reddening, initial mass function, starburst age, metallicity, and the escape fraction of ionizing photons, $f_{esc}$  \citep{steidel01,shapley06,siana07}.

\begin{figure}
\epsscale{1.0}
\plotone{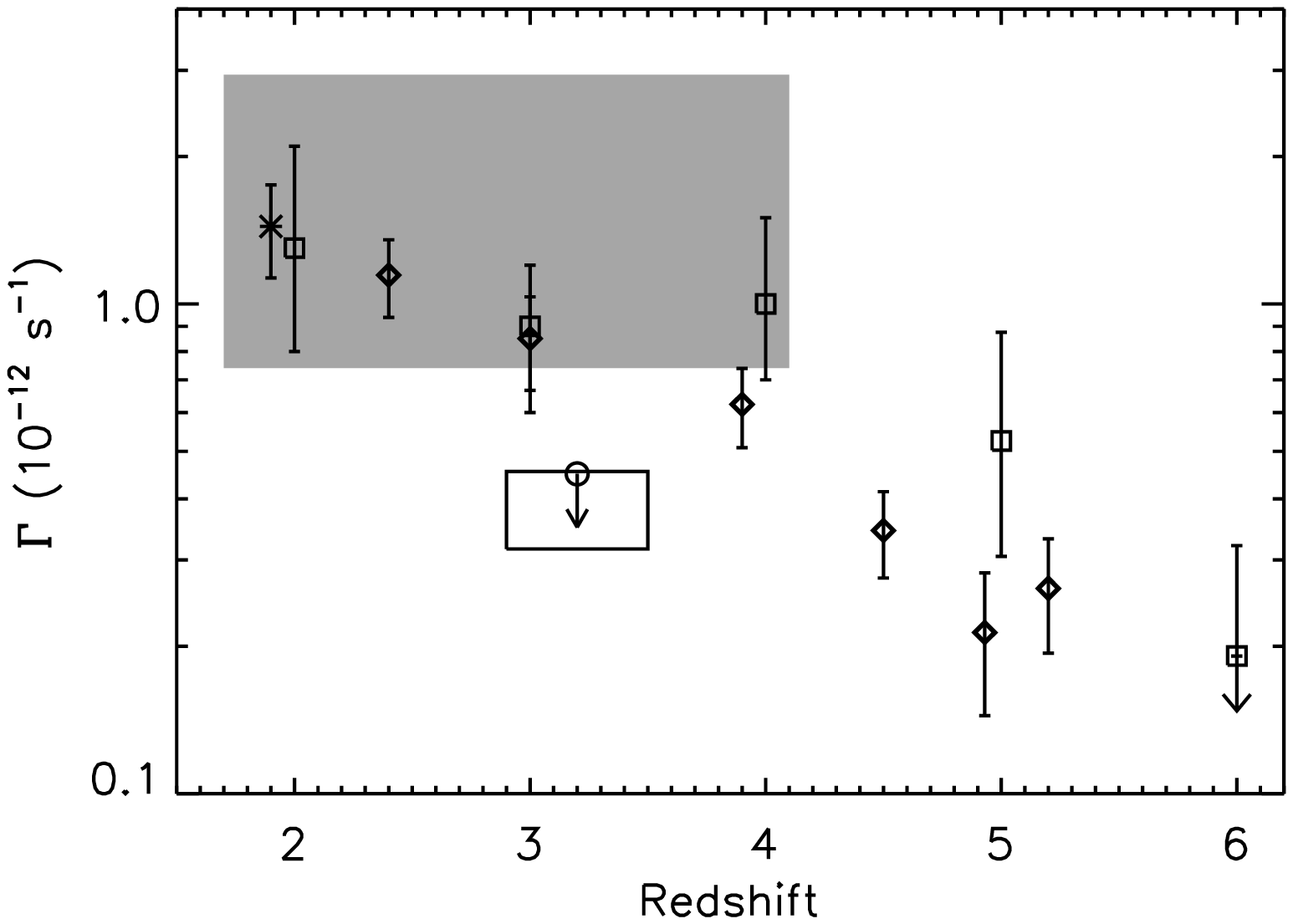}
\caption{HI photoionization rate per atom (in units of $1\times10^{-12}$ s$^{-1}$) versus redshift.  Estimates of the total photoionization rate of the IGM are plotted as asterisks \citep{tytler04}, diamonds \citep{mcdonald01}, and squares \citep{bolton05,bolton07}.  The shaded region is the photoionization rate determined by \citet{scott00} adjusted to our cosmology (decrease of 31\%) as stated in \citet{scott02}.  The open circle is the \citet{hunt04} determination of the QSO contribution at $z\sim3$.  The black square encompasses the redshift range and plausible limits of the QSO contribution to HI photoionization from our sample.  The lower and upper bounds are determined by integrating the QSO luminosity function to $R<22$ ($M_{1450}<-23.5$) and $R<25.5$ ($M_{1450}<-20$), respectively. \label{fig:ion_bkg}}
\end{figure}

It is possible, however, to address this question indirectly if the {\it total} ionizing background (QSOs+galaxies) is accurately determined.  Several groups have made these measurements by measuring the mean transmission of QSO UV flux through the Ly$\alpha$ forest \citep{mcdonald01,tytler04,bolton05,bolton07,becker07}, or by measuring the extent of the proximity effect \citep{carswell82} around high redshift QSOs \citep{scott00,scott02}.  Figure \ref{fig:ion_bkg} shows the current estimates for the total photoionization rate at high redshift, compared with our estimate of the contribution from QSOs at $z\sim3.2$.  The lower limit to $\Gamma$ is derived by integrating the QLF between $30<M_{1450}<23.5$, the range covered by our survey and SDSS.  The upper limit is derived by integrating the QLF 3.5 magnitudes fainter and assumes a 100\% escape fraction amongst these faint QSOs as well.  The \citet{scott00} value has been scaled (by the value given in \citet{scott02}) to our assumed cosmology.  The \citet{mcdonald01} values have also been scaled to the same cosmology using Eqn. 3 in their paper.  

Although the error bars are large, all of the {\it total} photoionization measurements are consistent and give a value of $\Gamma \sim 1.0\times10^{12}$ s$^{-1}$ at $z=3$.  The contribution from QSOs is less than half of this value.  Therefore, it is likely that star-forming galaxies' contribution to the HI photoionization rate is comparable to that of QSOs at $z\sim3.2$.  This is consistent with measurements by \citet{shull04} which examine the relative rates of HeII and HI photoionization within individual Ly$\alpha$ absorbers at $2.3<z<2.9$ to infer the spectral index of background radiation at each location.  They conclude that the spectral index varies greatly between absorbers, with significant contribution from ``soft'' sources that may be starburst galaxies or dust-attenuated AGN.  

Furthermore, recent measurements of the escape fraction, $f_{esc}$, of photoionizing radiation from Lyman Break Galaxies also suggest that star-formation may significantly contribute to the ionizing background at $z\sim3$.  \citet{steidel01} made a composite rest-frame ultraviolet spectrum of 29 LBGs and found that greater than 50\% of the photoionizing flux that is not absorbed by dust escapes into the IGM (ie. {\it relative} escape fraction $f_{esc,rel} > 0.5$).  Deeper spectra of 14 $z \sim 3$ LBGs give a smaller value of $f_{esc,rel} = 0.14$, but this still gives an ionizing radiation field $J_{900} \sim 2.6 \times 10^{-22}$ erg s$^{-1}$ cm$^{-2}$ Hz$^{-1}$, nearly twice the value of our upper limit from QSOs.  

\section{Summary}

We present our method of finding high redshift, $z>2.8$, QSOs by identifying a Lyman Break in the optical photometry, and ensuring red mid-IR ($[3.6]-[4.5]$) colors indicative of QSOs.  The use of only three optical filters allows a search over larger areas in the SWIRE fields as most of the area does not have coverage in four or more bands.  The use of only IRAC1 and IRAC2 channels is emphasized as these two bands are a factor of seven times more sensitive than IRAC3 and IRAC4.  

Spectroscopic follow-up of 10 $z\sim3$ ($U$-dropout) candidates confirms that all 10 are QSOs between $2.83<z<3.44$.  Spectroscopy of 10 $z\sim4$ ($g'$-dropout) candidates confirmed 7 QSOs with $3.48<z<3.88$, two galaxies at low redshift ($z=0.354$, $0.390$) and one unconfirmed redshift.  We place reliability estimates on our $z\sim3$ and $z\sim4$ samples of 100\% ($>69$\% $1\sigma$) and $70^{+16}_{-26}$\%, respectively.  Since we have not spectroscopically confirmed all of our candidates, we only use the more reliable $z\sim3$ sample for determining a luminosity function.  

By using detailed models which include variations in number and column density of line-of-sight HI absorbers, UV spectral slope, emission line equivalent width, redshift, observed magnitude, and photometric errors, we assess the completeness of the optical color selection.   Completeness near the center of the redshift range of our $U$-dropout selection is 85-90\%.  However, our completeness decreases significantly in our faintest magnitudes bins ($\sim 75$\%), due to the shallow depth of the $U$-band imaging.  

We find 100 $z \sim 3$ QSO candidates with $r'<22$ over 11.7 deg$^2$.  Through our models of completeness versus redshift, we derive effective volumes for each half-magnitude bin and compute the $z\sim3$ QSO luminosity function.  When combined with SDSS data, a least-squares fit to a double power-law gives a faint-end slope, $\beta = -1.62\pm0.19$, and location of the break at $M^*= -25.6$.  

Our binned QLF assumes that all of the QSO candidates are at $z=3.2$, which may skew the fitted parameters because of luminosity function evolution over our redshift range and Eddington Bias \citep{eddington13} due to large dispersions in the actual absolute magnitude distribution.  Therefore, we have performed a maximum likelihood fit of the {\it apparent} magnitude distribution of our sample with that inferred from a specific QLF over this redshift range.  Our results are slightly different, with a shallower faint-end slope, $\beta = -1.43 \pm 0.15$, and a somewhat fainter break at $M^*= -24.9$.  This fit is more accurate as it does not assume that all of the QSOs are at the same redshift.  

The fitted slope is consistent, within the errors, with values measured at low redshift ($0.5<z<2.0$), $\beta=-1.45$ \citep{richards05} and therefore does not require evolution in the faint end slope of the luminosity function.  Our QLF predicts significantly more faint QSOs than suggested with initial measurements at $z\sim3$ \citep{hunt04}.  Although it is difficult to tell with our limited spectroscopic sample, some of the difference between our faint-end slope and that of \citet{hunt04} may be attributed to an increasing number of narrow-line, moderately reddened AGN at fainter UV luminosities which were excluded from the H04 sample.  

The QLF exhibits some curvature at all magnitudes and, because of this, the parameters for a double power-law fit are degenerate.  That is, the position of the break, ($M^*$,$\Phi^*$), can be fit at different locations along the binned QLF, with appropriate changes in the bright and faint end slopes ($\alpha$,$\beta$).  This is especially true at high redshift, where the difference between the bright and faint end slopes appears to decrease.  Therefore, one must be careful when assigning physical significance to the measured values of these parameters when comparing to models.

The QSOs in our sample span the break in the luminosity function ($0.25L^* < L < 4.0L*$) and thus we measure the space density of QSOs that comprise the majority (55\%) of the QSO UV luminosity density at these redshifts.  When combined with the SDSS sample this percentage is more than $70$\%.  Therefore, large extrapolations are not required to estimate the effects of undetected QSOs.  The integrated UV luminosity density at $z \sim 3.2$ is $\epsilon_{1450} = 7.3\times10^{24}$ ergs s$^{-1}$ Hz$^{-1}$ $h$ Mpc$^{-3}$.  Using the scaling relation derived by \citet{haardt96}, we infer a maximum HI photoionization rate by QSOs, $\Gamma = 4.5\times10^{-13}$s$^{-1}$.  This is about 50\% of the {\it total} IGM HI photionization rate at $z=3$, requiring comparable ionizing flux from either starburst galaxies or redder AGN that lie outside our color criteria.

\acknowledgments

This work is based on observations made with the {\it Spitzer Space Telescope}, which is operated by the Jet Propulsion Laboratory, California Institute of Technology under a contract with NASA.  Support for this work, part of the {\it Spitzer Space Telescope} Legacy Science Program, was provided by NASA through an award issued by JPL/Caltech, under NASA contract 1407.

Based in part on observations obtained at the Hale Telescope, Palomar Observatory as part of a continuing collaboration between the California Institute of Technology, NASA/JPL, and Cornell University.  

Based in part on data made publically available through the Isaac Newton Groups' Wide Field Camera Survey Programme.  The Isaac Newton Telescope is operated on the island of La Palma by the Isaac Newton Group in the Spanish Observatorio del Canarias.

This publication makes use of data products from the Two Micron All Sky Survey, which is a joint project of the University of Massachusetts and the Infrared Processing and Analysis Center/California Institute of Technology, funded by the National Aeronautics and Space Administration and the National Science Foundation.

{\it Facilities:} \facility{Spitzer(IRAC,MIPS)}

\bibliography{apj-jour,all_ref}

\clearpage

\begin{deluxetable}{ccc}
\tabletypesize{\footnotesize}
\tablecaption{SWIRE IR Depths from \citet{swire_doc} \label{tab:ir_data}}
\tablewidth{0pt}
\tablehead{\colhead{Filter} & \colhead{Central Wavelength ($\mu m$)} & \colhead{Depth ($\mu Jy$, 5$\sigma$)}}
\startdata
IRAC1	&	3.6	&	6 \\
IRAC2 	&	4.5	&	7 \\
IRAC3	&	5.8	&	42\\
IRAC4	&	8.0 &	50\\
MIPS24	&  24.0 &  250\\
\enddata
\end{deluxetable}

\begin{deluxetable}{ccccc}
\tabletypesize{\footnotesize}
\tablecaption{WFS Optical Depths \label{tab:opt_data}}
\tablewidth{0pt}
\tablehead{\colhead{Filter} & \colhead{Central Wavelength (\AA)} & \colhead{Width (\AA)} & \colhead{$m_{AB}(Vega)$\tablenotemark{a}} & \colhead{Depth (Vega, 5$\sigma$)}}
\startdata
$U$	&	3560	&	600		&	0.78 	&	24.3	\\
$g'$&	4857	&	1400	&	-0.09	&	25.2	\\
$r'$&	6216	&	1380	&	0.15	&	24.5	\\
$i'$&	7671	&	1535	&	0.40	&	23.7	\\
$Z$	&	9100	&	1370	&	0.54	&	22.1	\\
\enddata
\tablenotetext{a}{$m_{AB}$(Vega) is the Vega to AB conversion factor where $m_{AB} = m_{Vega} + m_{AB}(Vega)$.}
\end{deluxetable}



\begin{deluxetable}{cc}
\tabletypesize{\footnotesize}
\tablecaption{The broadband optical/IR template combined with the \citet{telfer02}/\citet{vanden_berk01} composite spectra. \label{tab:template}}
\tablewidth{2in}
\tablehead{\colhead{$\lambda$ ($\mu$m)} & \colhead{$f_{\lambda}$}}
\startdata
   0.0302  &    2.250   \\
   0.0303  &    4.086   \\
   0.0304  &    6.034   \\
   0.0305  &    5.841   \\
   0.0306  &    6.000   \\
\enddata
\tablecomments{[The complete version of this table is in the electronic edition of the Journal.  The printed edition contains only a sample.]}
\end{deluxetable}


\begin{deluxetable}{c | c}
\tabletypesize{\footnotesize}
\tablecaption{Optical Selection Criteria. \label{tab:opt_sel}}
\tablewidth{0pt}
\tablehead{\colhead{Sample} & \colhead{Color Criteria (Vega)}}
\startdata
			&	$U-g'\geq 0.33$ \\
$z\sim3$		&	$g'-r'\leq 1.0$ \\
			&	$U-g'\geq 3.9\times(g'-r')-2.0$ \\
			&	\\
\cline{1-2}
\\
			&	$g'-r'\geq 1.241$ \\
$z\sim4$		&	$r'-i'\leq 1.146 $ \\
			&	$g'-r'\geq 2.178\times(r'-i')+0.09$ \\
			&	\\
\enddata
\end{deluxetable}

\clearpage
\LongTables
\begin{landscape}
\begin{deluxetable}{rcccccccccrrrrr}
\tabletypesize{\scriptsize}
\tablecaption{Optical/IR Photometry of the 100 $z\sim3$ QSO candidates.  Spectroscopic redshifts are given when available. $U$-band magnitudes in parentheses are 5$\sigma$ limits.  Two objects with spectroscopic confirmation were moved just outside of our color criteria (due to revised photometry) and are no longer part of our primary sample.  We list these at the end of the table. \label{tab:z3_cand}}
\tablewidth{0pt}
\tablehead{
\colhead{No.} & \colhead{Name} & \colhead{RA} & \colhead{Dec} & \colhead{$z_{spec}$} & \colhead{$U$} & \colhead{$g'$} & \colhead{$r'$} & \colhead{$i'$} & \colhead{$Z$} & \colhead{IRAC1} & \colhead{IRAC2} & \colhead{IRAC3} & \colhead{IRAC4} & \colhead{MIPS24} \\
\colhead{} & \colhead{} & \colhead{[Deg]} & \colhead{[Deg]} & \colhead{} &  \colhead{[Vega]} & \colhead{[Vega]} & \colhead{[Vega]} & \colhead{[Vega]} & \colhead{[Vega]} & \colhead{[$\mu$Jy]} & \colhead{[$\mu$Jy]} & \colhead{[$\mu$Jy]} & \colhead{[$\mu$Jy]} & \colhead{[$\mu$Jy]}
}
\startdata
   1  &  SWIRE\_J155907.09+550325.5  &  239.77954  &   55.05709  & \nodata  & 22.08  & 21.37  & 21.42  & 21.17  & 20.86  &    12  &    20  &    44  &    46  &  $<$250  \\
   2  &  SWIRE\_J160123.47+553750.3  &  240.34778  &   55.63064  & \nodata  & 22.75  & 22.19  & 21.62  & 21.14  & 21.34  &    20  &    20  & $<$42  & $<$50  &  $<$250  \\
   3  &  SWIRE\_J160153.49+542356.6  &  240.47287  &   54.39906  & \nodata  & 22.15  & 21.37  & 21.56  & 21.01  & 20.36  &    99  &   100  &   137  &   138  &     593  \\
   4  &  SWIRE\_J160223.85+552012.2  &  240.59938  &   55.33673  & \nodata  & 21.92  & 21.48  & 21.09  & 20.59  & 19.95  &    28  &    29  &    50  &    57  &     336  \\
   5  &  SWIRE\_J160318.38+552703.2  &  240.82657  &   55.45089  & \nodata  & 21.51  & 20.95  & 20.39  & 20.01  & 19.63  &    56  &    83  &   133  &   225  &     827  \\
   6  &  SWIRE\_J160345.78+542337.2  &  240.94077  &   54.39367  & \nodata  & 20.58  & 20.20  & 19.68  & 19.23  & 19.04  &    12  &    11  & $<$42  & $<$50  &  $<$250  \\
   7  &  SWIRE\_J160422.50+535454.9  &  241.09373  &   53.91525  & \nodata  & 22.01  & 21.23  & 20.60  & 20.18  & 20.03  &    26  &    35  &    60  &   118  &     473  \\
   8  &  SWIRE\_J160426.31+553446.9  &  241.10963  &   55.57969  & \nodata  & (23.30)  & 21.37  & 20.91  & 20.67  & 20.57  &    22  &    24  &    41  &    59  &  $<$250  \\
   9  &  SWIRE\_J160452.28+542758.2  &  241.21782  &   54.46617  &    2.98  & 22.17  & 21.15  & 21.30  & 21.13  & 20.86  &    28  &    34  & $<$42  &    67  &  $<$250  \\
  10  &  SWIRE\_J160520.52+552704.6  &  241.33548  &   55.45128  &    3.30  & 22.88  & 20.51  & 20.26  & 20.19  & 20.13  &    48  &    50  &    66  &    73  &     461  \\
  11  &  SWIRE\_J160528.14+560635.6  &  241.36725  &   56.10989  & \nodata  & 22.88  & 22.21  & 21.78  & 21.52  & 20.67  &    22  &    21  & $<$42  &    45  &     712  \\
  12  &  SWIRE\_J160606.33+550412.3  &  241.52638  &   55.07008  & \nodata  & 22.83  & 22.25  & 21.89  & 21.55  & 21.13  &    16  &    19  & $<$42  &    35  &  $<$250  \\
  13  &  SWIRE\_J160617.56+541649.5  &  241.57318  &   54.28042  & \nodata  & 23.60  & 22.00  & 21.21  & 20.74  & 21.19  &    32  &    37  &    59  &   116  &     290  \\
  14  &  SWIRE\_J160621.18+552532.7  &  241.58827  &   55.42574  & \nodata  & 23.12  & 21.59  & 21.11  & 20.68  & 20.21  &    53  &    61  &    76  &    78  &     382  \\
  15  &  SWIRE\_J160637.88+535008.4  &  241.65784  &   53.83568  &   2.943\tablenotemark{a}  & 20.63  & 19.79  & 19.47  & 18.95  & 18.75  &   109  &   169  &   312  &   739  &    2904  \\
  16  &  SWIRE\_J160654.19+554028.4  &  241.72578  &   55.67455  &    2.97  & 22.45  & 21.72  & 21.20  & 20.82  & 20.63  &    21  &    29  & $<$42  &    58  &     306  \\
  17  &  SWIRE\_J160656.10+535633.4  &  241.73373  &   53.94261  & \nodata  & (23.33)  & 21.53  & 21.12  & 20.66  & 20.57  &    32  &    33  & $<$42  &    68  &     276  \\
  18  &  SWIRE\_J160724.00+533615.2  &  241.85002  &   53.60422  & \nodata  & (23.25)  & 21.77  & 21.12  & 20.69  & 20.48  &    15  &    17  & $<$42  &    45  &     203  \\
  19  &  SWIRE\_J160733.94+554428.7  &  241.89142  &   55.74130  & \nodata  & 21.24  & 20.39  & 19.69  & 19.04  & 18.45  &   112  &   138  &   228  &   375  &     995  \\
  20  &  SWIRE\_J160754.39+533916.6  &  241.97661  &   53.65462  &    3.01  & 23.29  & 22.39  & 21.93  & 21.49  & 21.37  &    20  &    23  & $<$42  & $<$50  &     202  \\
  21  &  SWIRE\_J160758.67+543137.8  &  241.99446  &   54.52716  & \nodata  & 21.46  & 20.85  & 20.59  & 20.19  & 19.84  &    44  &    43  & $<$42  &    80  &  $<$250  \\
  22  &  SWIRE\_J160824.08+542003.7  &  242.10033  &   54.33435  & \nodata  & 22.98  & 22.39  & 21.98  & 21.71  & 21.04  &    17  &    16  & $<$42  & $<$50  &  $<$250  \\
  23  &  SWIRE\_J160850.55+545800.5  &  242.21063  &   54.96680  & \nodata  & (23.34)  & 22.60  & 21.92  & 21.52  & 21.45  &    11  &    11  & $<$42  & $<$50  &  $<$250  \\
  24  &  SWIRE\_J160907.33+543329.8  &  242.28056  &   54.55827  & \nodata  & 22.85  & 21.16  & 20.57  & 20.33  & 20.15  &    30  &    36  &    49  &   113  &     443  \\
  25  &  SWIRE\_J160917.26+553638.2  &  242.32191  &   55.61062  & \nodata  & 22.85  & 21.33  & 20.96  & 20.64  & 20.21  &    50  &    62  &    57  &    87  &     305  \\
  26  &  SWIRE\_J160947.86+552542.9  &  242.44943  &   55.42857  & \nodata  & 20.47  & 19.91  & 19.35  & 19.03  & 18.73  &    71  &    84  &    82  &   208  &     690  \\
  27  &  SWIRE\_J161008.08+552944.2  &  242.53366  &   55.49561  & \nodata  & 22.56  & 21.73  & 21.46  & 21.17  & 20.92  &    14  &    14  & $<$42  & $<$50  &  $<$250  \\
  28  &  SWIRE\_J161008.34+533254.9  &  242.53476  &   53.54857  &    3.20  & 22.81  & 21.51  & 20.99  & 20.18  & 19.78  &    63  &    66  &    82  &   150  &     494  \\
  29  &  SWIRE\_J161051.96+531004.4  &  242.71651  &   53.16788  & \nodata  & 22.79  & 21.82  & 21.09  & 20.56  & 20.15  &    32  &    37  &    58  &   110  &     417  \\
  30  &  SWIRE\_J161115.37+534029.2  &  242.81406  &   53.67478  & \nodata  & 22.89  & 22.28  & 21.77  & 20.75  & 20.05  &   121  &   117  &   132  &   135  &  $<$250  \\
  31  &  SWIRE\_J161128.36+553409.9  &  242.86815  &   55.56943  & \nodata  & (23.08)  & 22.35  & 21.70  & 21.34  & 21.27  &    15  &    24  & $<$42  & $<$50  &  $<$250  \\
  32  &  SWIRE\_J161128.47+535809.3  &  242.86862  &   53.96926  & \nodata  & (23.34)  & 22.24  & 21.94  & 21.46  & 21.72  &    22  &    38  &    58  &   190  &     754  \\
  33  &  SWIRE\_J161132.10+542312.0  &  242.88374  &   54.38667  & \nodata  & 22.86  & 21.23  & 20.58  & 20.43  & 20.26  &    48  &    56  &    70  &   131  &     299  \\
  34  &  SWIRE\_J161142.40+533104.6  &  242.92667  &   53.51794  &    3.06  & 22.42  & 21.35  & 21.04  & 20.77  & 20.51  &    37  &    43  &    57  &    86  &     302  \\
  35  &  SWIRE\_J161202.93+532346.9  &  243.01221  &   53.39636  & \nodata  & (23.12)  & 21.23  & 20.51  & 20.20  & 19.97  &    26  &    29  & $<$42  & $<$50  &     195  \\
  36  &  SWIRE\_J161251.98+534608.8  &  243.21658  &   53.76910  &    3.44  & 23.80  & 21.68  & 20.93  & 20.55  & 20.59  &    20  &    22  & $<$42  & $<$50  &  $<$250  \\
  37  &  SWIRE\_J161300.88+544629.6  &  243.25368  &   54.77489  & \nodata  & 23.27  & 21.75  & 21.74  & 21.09  & 21.73  &    27  &    38  & $<$42  &    76  &     166  \\
  38  &  SWIRE\_J161311.87+542403.7  &  243.29944  &   54.40102  & \nodata  & 22.65  & 22.18  & 21.83  & 21.46  & 21.27  &    29  &    32  &    53  &    43  &     237  \\
  39  &  SWIRE\_J161326.39+530923.0  &  243.35994  &   53.15638  &    2.83  & 20.34  & 19.88  & 19.47  & 19.25  & 19.04  &    68  &    94  &   124  &   217  &     574  \\
  40  &  SWIRE\_J161341.27+532956.4  &  243.42195  &   53.49900  & \nodata  & 22.79  & 21.95  & 21.24  & 21.05  & 21.26  &    23  &    31  & $<$42  &    78  &  $<$250  \\
  41  &  SWIRE\_J161430.84+555133.3  &  243.62848  &   55.85926  & \nodata  & 22.74  & 21.85  & 21.53  & 21.20  & 21.41  &    22  &    28  &    48  &    73  &     208  \\
  42  &  SWIRE\_J161433.14+533249.6  &  243.63808  &   53.54710  & \nodata  & 21.04  & 20.15  & 19.55  & 19.36  & 19.07  &   112  &   124  &   151  &   306  &    1068  \\
  43  &  SWIRE\_J161442.44+554614.2  &  243.67683  &   55.77060  & \nodata  & 22.86  & 21.68  & 21.12  & 20.76  & 20.48  &    19  &    22  & $<$42  &    56  &  $<$250  \\
  44  &  SWIRE\_J161446.30+555229.5  &  243.69290  &   55.87486  & \nodata  & (23.19)  & 22.24  & 21.53  & 20.86  & 20.76  &    28  &    38  & $<$42  &   107  &     308  \\
  45  &  SWIRE\_J161508.88+555514.6  &  243.78700  &   55.92073  & \nodata  & 20.55  & 19.71  & 19.29  & 18.92  & 18.78  &    55  &    75  &   122  &   301  &     924  \\
  46  &  SWIRE\_J161526.80+555217.4  &  243.86165  &   55.87150  & \nodata  & 22.50  & 21.96  & 21.33  & 20.66  & 20.19  &    67  &    78  &   117  &   237  &     810  \\
  47  &  SWIRE\_J161530.88+555247.1  &  243.87868  &   55.87976  & \nodata  & 22.91  & 22.37  & 21.97  & 21.49  & 20.75  &    58  &    58  &    60  & $<$50  &  $<$250  \\
  48  &  SWIRE\_J161549.80+540834.9  &  243.95749  &   54.14304  & \nodata  & (23.18)  & 22.25  & 21.84  & 21.66  & 22.05  &     9  &    11  & $<$42  & $<$50  &  $<$250  \\
  49  &  SWIRE\_J161626.05+535132.9  &  244.10854  &   53.85915  & \nodata  & (23.42)  & 21.48  & 20.82  & 20.29  & 19.93  &    43  &    45  &    63  &    98  &     286  \\
  50  &  SWIRE\_J161634.24+553528.2  &  244.14265  &   55.59116  & \nodata  & 20.14  & 19.68  & 19.29  & 19.01  & 18.86  &    55  &    66  &    89  &   161  &     705  \\
  51  &  SWIRE\_J161638.27+555701.4  &  244.15947  &   55.95039  & \nodata  & 21.60  & 20.82  & 20.42  & 19.93  & 19.80  &    46  &    62  &    72  &   153  &     616  \\
  52  &  SWIRE\_J161704.50+541200.4  &  244.26875  &   54.20012  & \nodata  & 23.24  & 22.13  & 21.48  & 21.15  & 21.08  &     8  &    10  & $<$42  & $<$50  &     265  \\
  53  &  SWIRE\_J161719.00+540154.3  &  244.32916  &   54.03175  & \nodata  & 22.66  & 22.05  & 21.42  & 21.17  & 20.98  &    28  &    40  &    45  &   107  &     356  \\
  54  &  SWIRE\_J161735.03+543830.8  &  244.39594  &   54.64189  & \nodata  & 23.66  & 22.42  & 21.84  & 21.64  & 21.32  &    16  &    21  & $<$42  &    44  &     381  \\
  55  &  SWIRE\_J161735.16+541405.9  &  244.39650  &   54.23497  & \nodata  & 22.68  & 20.08  & 19.57  & 19.22  & 19.11  &    56  &    64  &   117  &   185  &     717  \\
  56  &  SWIRE\_J161849.23+543658.3  &  244.70511  &   54.61620  & \nodata  & 22.37  & 21.91  & 21.53  & 21.28  & 21.13  &    15  &    17  & $<$42  &    38  &  $<$250  \\
  57  &  SWIRE\_J161936.10+541701.2  &  244.90041  &   54.28368  & \nodata  & 20.10  & 19.74  & 20.25  & 20.10  & 19.84  &   195  &   235  &   329  &   378  &    1212  \\
  58  &  SWIRE\_J161959.51+551453.8  &  244.99796  &   55.24828  & \nodata  & (23.29)  & 22.05  & 21.78  & 21.87  & 21.48  &    12  &    12  & $<$42  &    42  &   283  \\
  59  &  SWIRE\_J162004.22+545023.4  &  245.01758  &   54.83984  & \nodata  & 22.80  & 20.68  & 20.45  & 20.22  & 20.12  &    45  &    56  &   104  &   224  &     989  \\
  60  &  SWIRE\_J163032.07+405733.9  &  247.63361  &   40.95943  & \nodata  & 21.05  & 20.70  & 20.12  & 19.69  & 19.49  &    11  &    10  & $<$42  & $<$50  &  $<$250  \\
  61  &  SWIRE\_J163130.11+403555.6  &  247.87546  &   40.59879  & \nodata  & 22.43  & 21.43  & 21.94  & 20.92  & 20.29  &   113  &    99  &   117  &   110  &     650  \\
  62  &  SWIRE\_J163219.12+404637.2  &  248.07965  &   40.77701  & \nodata  & 20.26  & 19.78  & 19.46  & 19.14  & 19.11  &    76  &   103  &   148  &   291  &    1032  \\
  63  &  SWIRE\_J163259.11+401056.6  &  248.24628  &   40.18239  & \nodata  & 22.88  & 22.17  & 21.76  & 21.58  & 21.19  &    20  &    23  & $<$42  &    60  &     307  \\
  64  &  SWIRE\_J163340.14+404733.0  &  248.41725  &   40.79249  & \nodata  & 22.65  & 22.26  & 21.99  & 21.40  & 21.02  &    26  &    33  & $<$42  & $<$50  &  $<$250  \\
  65  &  SWIRE\_J163343.54+403739.5  &  248.43143  &   40.62764  & \nodata  & 20.91  & 20.34  & 19.71  & 19.31  & 19.14  &    36  &    34  & $<$42  & $<$50  &  $<$250  \\
  66  &  SWIRE\_J163357.79+400225.5  &  248.49081  &   40.04041  & \nodata  & (23.18)  & 21.44  & 20.83  & 20.56  & 20.57  &    34  &    37  &    48  &    56  &  $<$250  \\
  67  &  SWIRE\_J163359.28+410921.1  &  248.49701  &   41.15587  & \nodata  & 21.52  & 21.17  & 20.81  & 20.42  & 19.93  &    40  &    44  &    65  &   108  &  $<$250  \\
  68  &  SWIRE\_J163403.26+403845.1  &  248.51360  &   40.64585  & \nodata  & 22.81  & 22.17  & 21.54  & 20.93  & 20.91  &    20  &    19  & $<$42  & $<$50  &  $<$250  \\
  69  &  SWIRE\_J163413.96+412028.3  &  248.55818  &   41.34119  & \nodata  & 22.44  & 21.41  & 20.92  & 20.54  & 20.53  &    20  &    26  &    48  &    98  &     376  \\
  70  &  SWIRE\_J163417.97+410531.9  &  248.57487  &   41.09219  & \nodata  & 21.12  & 20.25  & 20.17  & 19.60  & 19.48  &    63  &    73  &   107  &   201  &     606  \\
  71  &  SWIRE\_J163423.23+400244.0  &  248.59680  &   40.04555  & \nodata  & 20.89  & 20.56  & 20.04  & 19.52  & 19.40  &    47  &    69  &   113  &   252  &    1080  \\
  72  &  SWIRE\_J163511.20+404335.4  &  248.79665  &   40.72651  & \nodata  & 21.62  & 21.04  & 20.64  & 19.94  & 19.75  &    39  &    41  &    43  &    86  &     180  \\
  73  &  SWIRE\_J163527.24+395907.4  &  248.86349  &   39.98538  & \nodata  & 21.89  & 21.33  & 20.85  & 20.45  & 20.25  &    40  &    47  &    74  &   171  &     590  \\
  74  &  SWIRE\_J163536.67+412338.7  &  248.90279  &   41.39407  & \nodata  & 22.65  & 22.02  & 21.77  & 21.22  & 21.54  &     7  &    12  & $<$42  & $<$50  &  $<$250  \\
  75  &  SWIRE\_J163537.37+414904.2  &  248.90570  &   41.81784  & \nodata  & 22.23  & 21.00  & 20.55  & 20.17  & 20.14  &    29  &    36  &    63  &   126  &     415  \\
  76  &  SWIRE\_J163553.80+412641.9  &  248.97417  &   41.44498  & \nodata  & 22.90  & 21.96  & 21.31  & 21.03  & 20.60  &    25  &    32  &    53  &    87  &  $<$250  \\
  77  &  SWIRE\_J163604.98+410307.7  &  249.02077  &   41.05215  & \nodata  & 22.26  & 21.82  & 21.21  & 20.71  & 20.57  &    13  &    15  & $<$42  & $<$50  &  $<$250  \\
  78  &  SWIRE\_J163627.59+405153.2  &  249.11494  &   40.86479  & \nodata  & 22.54  & 21.16  & 20.65  & 20.18  & 19.99  &    28  &    38  &    37  &   137  &     653  \\
  79  &  SWIRE\_J163627.66+404218.8  &  249.11525  &   40.70521  & \nodata  & 21.68  & 21.20  & 20.63  & 20.19  & 19.98  &    14  &    14  & $<$42  & $<$50  &  $<$250  \\
  80  &  SWIRE\_J163631.32+412904.7  &  249.13049  &   41.48464  & \nodata  & 22.14  & 21.71  & 21.11  & 20.21  & 19.78  &   208  &   313  &   332  &   451  &    1554  \\
  81  &  SWIRE\_J163657.12+412850.0  &  249.23801  &   41.48055  & \nodata  & 21.90  & 21.55  & 21.17  & 20.78  & 20.33  &    17  &    26  & $<$42  &    58  &  $<$250  \\
  82  &  SWIRE\_J163711.53+415912.2  &  249.29805  &   41.98671  & \nodata  & (23.38)  & 22.51  & 21.96  & 21.59  & 21.54  &     7  &    10  & $<$42  & $<$50  &  $<$250  \\
  83  &  SWIRE\_J163723.50+414757.5  &  249.34790  &   41.79930  & \nodata  & 22.86  & 21.86  & 21.46  & 21.09  & 21.15  &    19  &    27  &    69  & $<$50  &  $<$250  \\
  84  &  SWIRE\_J163733.22+413116.4  &  249.38840  &   41.52121  & \nodata  & (23.30)  & 21.88  & 21.80  & 21.15  & 21.03  &    14  &    19  & $<$42  &    37  &  $<$250  \\
  85  &  SWIRE\_J163744.75+414245.6  &  249.43646  &   41.71266  & \nodata  & 21.17  & 20.64  & 20.10  & 19.52  & 19.46  &    32  &    38  &    71  &   151  &     634  \\
  86  &  SWIRE\_J163822.19+403650.9  &  249.59245  &   40.61413  & \nodata  & 21.94  & 20.97  & 20.66  & 20.40  & 20.40  &    28  &    28  & $<$42  &    52  &  $<$250  \\
  87  &  SWIRE\_J163834.42+410014.9  &  249.64342  &   41.00415  & \nodata  & 22.44  & 21.75  & 21.62  & 21.65  & 21.29  &    13  &    14  &    35  &    52  &     264  \\
  88  &  SWIRE\_J163852.28+410923.7  &  249.71785  &   41.15657  & \nodata  & 22.43  & 22.06  & 21.79  & 20.89  & 20.62  &    56  &    85  &   108  &   169  &     476  \\
  89  &  SWIRE\_J163853.36+421433.2  &  249.72234  &   42.24255  & \nodata  & 20.95  & 20.56  & 20.17  & 20.05  & 20.03  &    95  &   133  &   209  &   309  &    1023  \\
  90  &  SWIRE\_J163916.07+414823.7  &  249.81694  &   41.80657  & \nodata  & (23.27)  & 22.06  & 21.32  & 21.00  & 20.61  &    18  &    23  & $<$42  & $<$50  &  $<$250  \\
  91  &  SWIRE\_J163918.81+412206.3  &  249.82837  &   41.36841  & \nodata  & (23.43)  & 20.89  & 20.36  & 19.88  & 19.85  &    43  &    52  &    84  &   176  &     441  \\
  92  &  SWIRE\_J163920.04+421745.8  &  249.83350  &   42.29605  & \nodata  & 22.80  & 21.83  & 21.31  & 21.10  & 20.84  &    32  &    35  & $<$42  & $<$50  &  $<$250  \\
  93  &  SWIRE\_J163940.71+403140.7  &  249.91962  &   40.52798  & \nodata  & 22.06  & 21.39  & 20.79  & 20.73  & 20.25  &    27  &    32  &    51  &    69  &  $<$250  \\
  94  &  SWIRE\_J163941.78+403909.8  &  249.92409  &   40.65273  & \nodata  & 22.53  & 22.19  & 21.70  & 21.27  & 20.98  &     8  &    11  & $<$42  & $<$50  &  $<$250  \\
  95  &  SWIRE\_J163955.93+410631.8  &  249.98306  &   41.10884  & \nodata  & (23.18)  & 22.27  & 21.80  & 21.22  & 20.91  &    20  &    33  &    49  &   131  &     527  \\
  96  &  SWIRE\_J163956.76+404550.7  &  249.98650  &   40.76407  & \nodata  & (23.16)  & 21.68  & 21.31  & 99.00  & 99.00  &    36  &    39  &    70  &   100  &  $<$250  \\
  97  &  SWIRE\_J164022.80+411548.2  &  250.09499  &   41.26338  &   3.064\tablenotemark{a}  & 20.28  & 19.72  & 19.20  & 18.88  & 18.76  &   100  &   126  &   187  &   350  &    1021  \\
  98  &  SWIRE\_J164036.64+414916.2  &  250.15268  &   41.82116  & \nodata  & 20.81  & 20.46  & 20.28  & 20.02  & 19.69  &    39  &    51  &    99  &   128  &     380  \\
  99  &  SWIRE\_J164127.88+412636.4  &  250.36617  &   41.44345  & \nodata  & 22.34  & 21.71  & 21.26  & 21.00  & 20.62  &    23  &    26  &    52  &    60  &     160  \\
 100  &  SWIRE\_J164229.64+405354.3  &  250.62350  &   40.89842  & \nodata  & (23.08)  & 22.59  & 21.98  & 21.65  & 21.19  &    56  &    63  &    55  &    64  &     285  \\
\hline
 101\tablenotemark{b}  &  SWIRE\_J160755.85+534020.4   &  241.98271  &   53.67234  & 3.07  & 22.46 & 22.23 & 21.51 & 21.27 & 21.02 & 22 & 27 & $<$42 & 54 & 185 \\
 102\tablenotemark{b}  &  SWIRE\_J161046.25+532540.5   &  242.69270  &   53.42791  & 2.98  & 21.41 & 21.11 & 20.68 & 20.51 & 20.71 & 36 & 35 & 54 & 46 & 241 \\
\enddata
\tablecomments{Typical uncertainties are $\sim0.04$ mags in the optical and $\sim10$\%\ in infrared fluxes.}
\tablenotetext{a}{$z_{spec}$ from SDSS.}
\tablenotetext{b}{Revised photometry moved object outside of color criteria.}
\end{deluxetable}
\clearpage
\end{landscape}

\clearpage
\begin{landscape}
\begin{deluxetable}{rcccccccccrrrrr}
\tabletypesize{\scriptsize}
\tablecaption{Redshifts and Optical/IR Photometry of the seven spectroscopically confirmed $z\sim4$ ($g'$-dropout) QSOs and two SDSS QSOs in our sample.  $U$-band magnitudes in parentheses are 5$\sigma$ limits.\label{tab:z4_spec}}
\tablewidth{0pt}
\tablehead{
\colhead{No.} & \colhead{Name} & \colhead{RA} & \colhead{Dec} & \colhead{$z_{spec}$} & \colhead{$U$} & \colhead{$g'$} & \colhead{$r'$} & \colhead{$i'$} & \colhead{$Z$} & \colhead{IRAC1} & \colhead{IRAC2} & \colhead{IRAC3} & \colhead{IRAC4} & \colhead{MIPS24} \\
\colhead{} & \colhead{} & \colhead{[Deg]} & \colhead{[Deg]} & \colhead{} &  \colhead{[Vega]} & \colhead{[Vega]} & \colhead{[Vega]} & \colhead{[Vega]} & \colhead{[Vega]} & \colhead{[$\mu$Jy]} & \colhead{[$\mu$Jy]} & \colhead{[$\mu$Jy]} & \colhead{[$\mu$Jy]} & \colhead{[$\mu$Jy]}
}
\startdata
   1  &  SWIRE\_J160532.15+542631.0  &  241.38394  &   54.44194  &    3.82  & 23.13  & 22.24  & 20.90  & 20.74  & 20.25  &    32  &    36  &    65  &   106  &     387  \\
   2  &  SWIRE\_J160705.15+533558.7  &  241.77147  &   53.59965  &   3.653\tablenotemark{a}  & (22.82)  & 19.47  & 18.18  & 17.89  & 17.55  &   374  &   412  &   665  &  1392  &    5727  \\
   3  &  SWIRE\_J160907.51+535028.0  &  242.28130  &   53.84110  &    3.62  & (22.73)  & 22.75  & 21.19  & 20.69  & 20.33  &    44  &    42  &    48  & $<$50  &     231  \\
   4  &  SWIRE\_J160934.11+550015.2  &  242.39211  &   55.00421  &    3.52  & 22.44  & 21.30  & 19.79  & 20.18  & 19.85  &    39  &    40  &    40  &    60  &  $<$250  \\
   5  &  SWIRE\_J161143.22+553157.5  &  242.93008  &   55.53263  &    3.58  & (23.04)  & 21.30  & 19.99  & 19.72  & 19.43  &    73  &    81  &   144  &   278  &    1368  \\
   6  &  SWIRE\_J161243.17+535827.4  &  243.17989  &   53.97428  &    3.83  & (23.34)  & 22.49  & 21.02  & 20.80  & 20.73  &    22  &    21  &    36  & $<$50  &  $<$250  \\
   7  &  SWIRE\_J161557.43+545915.8  &  243.98929  &   54.98772  &   3.880  & (23.29)  & 23.96  & 21.77  & 21.20  & 21.34  &    36  &    30  & $<$42  &    55  &  $<$250  \\
   8  &  SWIRE\_J164326.24+410343.4  &  250.85934  &   41.06205  &   3.873\tablenotemark{a}  & (23.11)  & 21.49  & 20.04  & 19.81  & 19.80  &    46  &    39  &    39  &    57  &     289  \\
\hline
   9\tablenotemark{b}  &  SWIRE\_J104350.94+583029.3  &  160.96227  &   58.50813  &   3.70   & \nodata & 20.95 & 19.43 & 18.97 & \nodata & 105 & 116 & 189 & 382 & 1945 \\
\hline
  10\tablenotemark{c}  &  SWIRE\_J160543.05+535829.2                     &  241.42938  &   53.97477  &   0.354  & (23.53) & 23.22 & 21.48 & 20.70 & 20.12 & 27        *    21 &  $<42$ & $<50$ & $<$250 \\
  11\tablenotemark{c}  &  SWIRE\_J161159.69+544211.3                     &  242.99872  &   54.70313 &   0.390  & 22.86 & 22.55 & 20.72 & 20.55 & 19.99 & 57 & 57 & $<$42 & 151 & $<$250 \\ 
  12\tablenotemark{bd}  &  SWIRE\_J105002.88+574720.0                     &  162.51199  &   57.78890 &  ? & \nodata & 21.55 & 20.05 & 19.73 & \nodata & 215 & 391 & 677 & 1220 & 3519 \\  
\enddata
\tablecomments{Typical uncertainties are $\sim0.04$ mags in the optical and $\sim10$\%\ in infrared fluxes.}
\tablenotetext{a}{ $z_{spec}$ from SDSS.}
\tablenotetext{b}{ Selected in the SWIRE Lockman field.}
\tablenotetext{c}{ Low redshift interlopers.}
\tablenotetext{d}{ Could not see emission lines for redshift confirmation.}
\end{deluxetable}
\clearpage
\end{landscape}

\begin{deluxetable}{lcc}
\tabletypesize{\footnotesize}
\tablecaption{Spectral parameter attribute for our QSO SED modeling. \label{tab:sim_params}}
\tablewidth{0pt}
\tablehead{\colhead{Parameter} & \colhead{Mean} & \colhead{$\sigma$}}
\startdata
$\alpha_{\nu}$				&	-0.46	&	0.3 \\
$W$(Ly$\alpha$+NV)(\AA)	&	101.7	& 	25. \\
\enddata
\end{deluxetable}

\begin{deluxetable}{lccccc}
\tabletypesize{\footnotesize}
\tablecaption{Parameters for the four column density ranges used in the Ly$\alpha$ Forest simulations.\label{tab:laf_params}}
\tablewidth{0pt}
\tablehead{\colhead{Name} & \colhead{log($N_{HI} $(cm$^2$))} & \colhead{$N_0$} & \colhead{$\kappa$} & \colhead{$\gamma$} & \colhead{b(km/s)}}
\startdata
Ly$\alpha$ Forest \#1\tablenotemark{a}	& 12-14		&	181.36	&	1.46	&	1.29	&	30	\\
Ly$\alpha$ Forest \#2\tablenotemark{a} 	& 14-17.2 	&	1.297	&	1.46	&	3.10	&	30	\\
Lyman Limit Systems\tablenotemark{b}		& 17.2-20	&	0.27	&	1.50	&	1.55	&	70 	\\
Damped Ly$\alpha$ Systems\tablenotemark{c} & 20-22	&	0.055	&	1.78	&	1.11	&	70	\\
\enddata
\tablenotetext{a}{\citet{kim97}}
\tablenotetext{b}{\citet{storrie-lombardi94}}
\tablenotetext{c}{\citet{storrie-lombardi00}}
\end{deluxetable}

\begin{deluxetable}{ccccccc}
\tabletypesize{\footnotesize}
\tablecaption{Tabulated Luminosity Function. \label{tab:tab_qlf}}
\tablewidth{0pt}
\tablehead{\colhead{$r'$} & \colhead{M(1450\AA)} & \colhead{n} & \colhead{Completeness} & \colhead{$z_{eff}$} & \colhead{$V_{eff}$} & \colhead{$\Phi$} \\
\colhead{Vega} & \colhead{AB} & \colhead{} & \colhead{} & \colhead{} & \colhead{($10^7$ Mpc$^3$)} & \colhead{($10^7$ mag$^{-1}$ Mpc$^{-3}$)}}
\startdata
18.25 & -27.11 &    0   &   1.00    &   3.22    &   7.87    &   $<0.47$ \\
18.75 & -26.61 &    0   &   1.00    &   3.22    &   7.87    &   $<0.47$ \\
19.25 & -26.11 &	7	&	1.00	&	3.22	&	7.87	&	1.78$^{+1.0}_{-0.7}$  \\
19.75 & -25.61 &	5	&	1.00	&	3.22	&	7.87	&   1.27$^{+0.9}_{-0.5}$  \\
20.25 & -25.11 &	12	&	0.99	&	3.22	&	7.83	& 	3.07$^{+1.2}_{-0.9}$  \\
20.75 & -24.61 &	20	&	0.98	&	3.21	&	7.73	& 	5.38$^{+1.5}_{-1.2}$  \\
21.25 & -24.11 &	25	&	0.86	&	3.18	&	6.77	& 	7.38$^{+1.8}_{-1.5}$  \\
21.75 & -23.61 &	31	&	0.74	&	3.15	&	5.82	& 	10.66$^{+2.3}_{-1.9}$  \\
\enddata
\end{deluxetable}

\clearpage

\begin{deluxetable}{lcccccc}
\tabletypesize{\footnotesize}
\tablecaption{Parameters for double power-law luminosity function.  The binned QLF assumes that all QSOs are at $z=3.2$, whereas the Maximum Likelihood values are from fits to the apparent magnitude distribution.  Values in parentheses denote $1\sigma$ errors and values in square brackets denote fixed parameters.  The maximimum-likelihood fits that include QLF evolution (both PDE and PLE) are not significantly different from the no evolution parameters.  Listed at the bottom are estimates from previous work. \label{tab:qlf_params}}
\tablewidth{0pt}
\tablehead{\colhead{Data} & \colhead{$\alpha$} & \colhead{$\beta$} & \colhead{$M^*_{1450}$ (AB)} & $\Phi^*$ (10$^{-7}$ mag$^{-1}$ Mpc$^{-3}$)  &  \colhead{$\chi ^2$} & \colhead{$\nu$\tablenotemark{a}}}
\startdata
{\it Binned QLF} \\
\\
SWIRE only						& -3.53 (2.9)  & -1.66 (0.88) & -25.8 (2.3) & 3.5  (11.0) & 2.4  & 3 \\
SWIRE only (fixed $\alpha$)     & [-2.85]      & -1.29 (0.85) & -24.8 (1.1) & 10.4 (12.0) & 2.5  & 4 \\
SWIRE+SDSS                      & -3.47 (0.58) & -1.98 (0.17) & -27.1 (0.7) & 0.57 (0.65) & 9.9  & 12 \\
SWIRE+SDSS (fixed $\alpha$) 	& [-2.85]      & -1.62 (0.19) & -25.6 (0.3) & 4.53 (2.0)  & 11.3 & 13 \\
\hline
{\it Maximum Likelihood Fit} \\
\\
SWIRE only (fixed $\alpha$, no evol)     & [-2.85] & -1.26 (0.21) & -25.0 (0.30) & 9.0 (3.0) \\
	       (PDE $(1+z)^{-3}$)			 & [-2.85] & -1.22 (0.22) & -25.0 (0.29) & 9.3 (2.9)  \\
		   (PLE )						 & [-2.85] & -1.25 (0.21) & -25.0 (0.30) & 9.2 (2.9)  \\
SWIRE+SDSS (fixed $\alpha$, no evol)     & [-2.85] & -1.43 (0.15) & -24.9 (0.15) & 8.5 (2.0) \\
           (PDE $(1+z)^{-3}$)			 & [-2.85] & -1.41 (0.15) & -24.9 (0.15) & 8.6 (1.9) \\
		   (PLE )						 & [-2.85] & -1.42 (0.15) & -24.9 (0.15) & 8.6 (1.8) \\
\hline
{\it Previous Studies} \\
\\
Hunt et al. (2004)                       & -4.56 (0.51)  & -1.24 (0.07) & -26.7 & 2.4 \\
Pei (1995)								 & -3.52 (0.11)  & -1.64 (0.18) & -25.8 (0.25) & 6.1 (2.5)\\
Bongiorno et al. (2007)					 & -3.0    & -1.38        & -25.7    & 9.8 \\
\enddata
\tablenotetext{a}{Degrees of freedom in $\chi^2$ fit.}
\end{deluxetable}

\end{document}